\documentclass[a4paper]{article}
\usepackage[T1]{fontenc}
\usepackage[utf8]{inputenc}
\DeclareUnicodeCharacter{0301}{\'{e}}
\usepackage[english,brazilian]{babel}
\usepackage{graphicx}

\usepackage{hyperref}
\hypersetup{
  setpagesize  = false,
  colorlinks   = true,    
  urlcolor     = blue,    
  linkcolor    = black,   
  citecolor    = black    
}
\usepackage{authblk}
\usepackage{geometry}
\usepackage{setspace}
\usepackage{verbatim}
\usepackage[utf8]{inputenc}
\usepackage{amsmath}
\usepackage{amssymb}
\usepackage{gensymb}
\usepackage{verbatim} 
\usepackage{ragged2e} 
\usepackage[english,brazilian]{babel}

\usepackage{cancel}
\usepackage{csquotes}

\usepackage[backend=biber]{biblatex}
\newcommand{\keywordsenglishname}{Keywords}

\renewenvironment{abstract}{%
        \begin{center}
	\begin{minipage}{14cm}
	{\textbf{\abstractname:}}
}{
        \end{minipage}
	\end{center}
}
\newenvironment{abstractinenglish}{
        \def\abstractname{\abstractinenglishname}
	\begin{abstract}
}{
        \end{abstract}
}

\newenvironment{keywordsenglish}{
        \def\abstractname{\emph{\keywordsenglishname}}
	\begin{abstract}
}{
        \end{abstract}
}

\title {Baryon Acoustic Oscillations from galaxy surveys\\[1ex] \large Baryon Acoustic Oscillations from galaxy surveys}
\author{Paula Ferreira\href{https://orcid.org/0000-0002-7540-040X}{\includegraphics[scale=0.04]{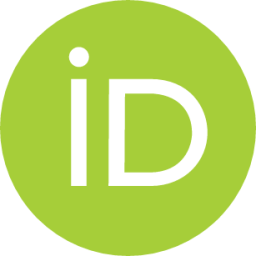}} \thanks{Endereço de correspondência: psfer@pos.if.ufrj.br} $^1$ }

\author{Ribamar R. R. Reis\href{https://orcid.org/0000-0003-1339-2106}{\includegraphics[scale=0.04]{orcid.png}} \thanks{Endereço de correspondência: ribamar@if.ufrj.br} $^{1,2}$ }
\affil{$^1$Instituto de Física, Universidade Federal do Rio de Janeiro, Rio de Janeiro, RJ, Brasil.}
\affil{$^{2}$Observatório do Valongo, Universidade Federal do Rio de Janeiro, Rio de Janeiro, RJ, Brasil.}
\date{}

\usepackage{fancyhdr}
\begin{document}

\maketitle
\vspace{6pt}

\begin{abstractinenglish}
We conducted a review of the fundamental aspects of describing and detecting the Baryon Acoustic Oscillation (BAO) feature in galaxy surveys, emphasizing the optimal tools for constraining this probe based on the type of observation. Additionally, we included new results with two spectroscopic datasets to determine the best-fit model for the power spectrum, $P(k)$. Using the framework described in a previous analysis, we applied this to a different sub-sample of the BOSS survey, specifically galaxies with redshifts $0.3<z<0.65$. We also examined the eBOSS dataset with redshifts $0.6<z<1.0$, adjusting the number of parameters in the traditional polynomial fit to account for the higher redshift range.
Our results showed that the dilation scale parameter $\alpha$ derived from the BOSS dataset had smaller error bars compared to the eBOSS dataset, attributable to the larger number of luminous red galaxies (LRGs) in the BOSS sample. We also compared our findings with other surveys such as WiggleZ, DES Y6, and DESI III, noting that photometric surveys typically yield larger error bars due to their lower precision. The DESI III results were in good agreement with ours within $1\sigma$, with most bins close to unity.
The variation of $\alpha$ with respect to the redshift is an unresolved issue in the field, appearing in both three-dimensional and angular tomographic analyses. 

\emph{}
\end{abstractinenglish}

\begin{keywordsenglish}
\emph{Observational Cosmology, BAO, LSS, galaxy surveys} 
\end{keywordsenglish}
\selectlanguage{english}

\section{Introduction}
 The cosmological standard model states that the universe was hot and dense in the past and has been expanding since then. During this early epoch, matter was completely ionised and tightly coupled to radiation, which is relativistic matter. As the Universe expanded and cooled down, matter arranged itself in neutral atoms in the so-called recombination process and decoupled from radiation.

The imprinted pattern left by the sound waves in the baryon-photon fluid on the last scattered photons is observed today as the Cosmic Microwave Background (CMB). Such waves are characterised by compression imposed gravitationally by baryons and expansion due to the pressure carried by photons that, due to their high temperatures in the primordial universe, acted like a fluid. Only after the decoupling from radiation could the baryonic matter fall into the gravitational potential of Cold Dark Matter (CDM), and the structures continued to grow to form today's observable universe with clusters of galaxies. The last known phase of evolution is the accelerating expansion due to the domination of dark energy ($\Lambda$) whose equation of state is a constant. The $\Lambda$CDM model describes the evolution of the universe from a hot dense period to the cooler state observed today. 

Before decoupling, small perturbations, due to quantum fluctuations amplified by inflation, travelled through the baryon-photon fluid as sound waves. After decoupling, the sound speed in the baryon fluid effectively went to zero due to negligible pressure, and the wavefronts were frozen as a pattern in the cosmic structure. The maximum distance travelled by these waves is the sound horizon $r_s$. The sound horizon at the drag epoch is called $r_s$, the scale on which the pressure of photons no longer overcomes the baryon gravitational instability. If we consider the $\Lambda\text{CDM}$ model with the inflationary scenario, we have $r_s \approx 150 $ Mpc. This is the Baryon Acoustic Oscillations (BAO) scale, where the photon instability can no longer deter the baryons gravitational force \cite{peebles1970primeval}, \cite{sunyaev1970small}. 

Two statistical methods to detect the BAO scale are the two-point correlation function and its Fourier transform, the power spectrum \parencite{peebles1973statistical}. The two-point correlation function measures the excess probability of finding a pair of galaxies separated by a given distance, while the power spectrum assesses the contributions of the energy density inhomogeneities on different scales \parencite{eisenstein1998baryonic}. The BAO scale appears as a bump in the two-point correlation function and as a series of wiggles in the matter power spectrum. 

The first detection from a relatively large survey was made by \cite{eisenstein2005detection}. The data comprised the first sample of Luminous Red Galaxies (LRG) from the Sloan Digital Sky Server (SDSS). The detection shown was the first evidence using the two-point correlation function, and the result indicated the position of the BAO peak at $z = 0.35$, the effective redshift of the sample. At the same time, the 2dF Galaxy Redshift Survey (2dFGRS) \cite{cole20052df} found the BAO signal using the power spectrum, and they showed the results to constrain the matter content in the universe using this type of probe for the first time.

\cite{tegmark2006cosmological} used the power spectra of the same data set from \cite{eisenstein2005detection} and compared the findings to the Wilkinson Microwave Anisotropy Probe (WMAP) \cite{hinshaw2007three}, which detected the acoustic scale through the Cosmic Microwave Background (CMB) power spectra. The LRG proved to agree with some of the WMAP's cosmological constraints. It is important to note that, although the power spectrum is the Fourier transform of the correlation function and would carry the same information in theory, this is not exactly the case for the observations since we cannot access all scales.

The current state of the art in CMB measurements is the successor of the WMAP, the Planck satellite \cite{collaboration2018planck}. This measurement combined with the most precise measurements of BAO using LRG from SDSS DR12 \cite{ross2017clustering}, \cite{vargas2018clustering}, and \cite{beutler2017clustering}; each with effective redshifts 0.38, 0.51, and 0.61. 

The matter power spectrum estimator was proposed by \cite{feldman1993power}, it uses the galaxy density field which is a function of the galaxy's and random catalog's number density.  This method relies on the weighted galaxy fluctuation field $F(\vec{r})$. To optimise the power spectrum estimator, a weight function $w(\vec{r})$ is introduced. These weights are constructed to get the least biased mean number density $\bar{n}(\vec{r})$.

An error term appears in the power spectrum estimation from \cite{feldman1993power}, the shot noise. The pattern expected to represent the BAO echo will only be detected if there is a sufficient number of targets. As the number of galaxies increases, the echo becomes clearer. The shot noise error decreases as the number of galaxies increases, making it sensitive to the volume of the survey.

This work covers various aspects of finding the BAO. We divided in a reviewing part and new results from a different sub-sample never used before. The first consists of 993,228 galaxies with a redshift range of $0.3<z<0.65$, while the second has $174,816$ galaxies with $0.6<z<1.0$. We provide the steps to obtain the galaxy power spectrum using the SDSS-III DR12 and DR16 catalogs. We performed the analysis with $\texttt{nbodykit}$'s \cite{hand2018nbodykit} open source code to compute the power spectrum. The BAO distance is obtained with \cite{anderson2014clustering} similar methodology, in which we find the scale parameter $\alpha$ and compare it with those found in the different redshift bins from other publications.

The study is organised as follows. In section \ref{sec:theo}, we introduce the equations that describe the BAO from the $\Lambda$CDM model. Next, in section \ref{sec:est} we describe some observables and estimators of the BAO feature. In section \ref{sec:mocks}, we explain the reasons and uses of mock catalogs.  Then, we give the details of the methods and the results to compute P(k) with BOSS and eBOSS in section \ref{sec:method} and \ref{sec:results}, respectively. Finally, we conclude in section \ref{sec:conclusion}.

\section{Theoretical background}\label{sec:theo}

\begin{figure}
    \centering
    \includegraphics[width=.8\textwidth]{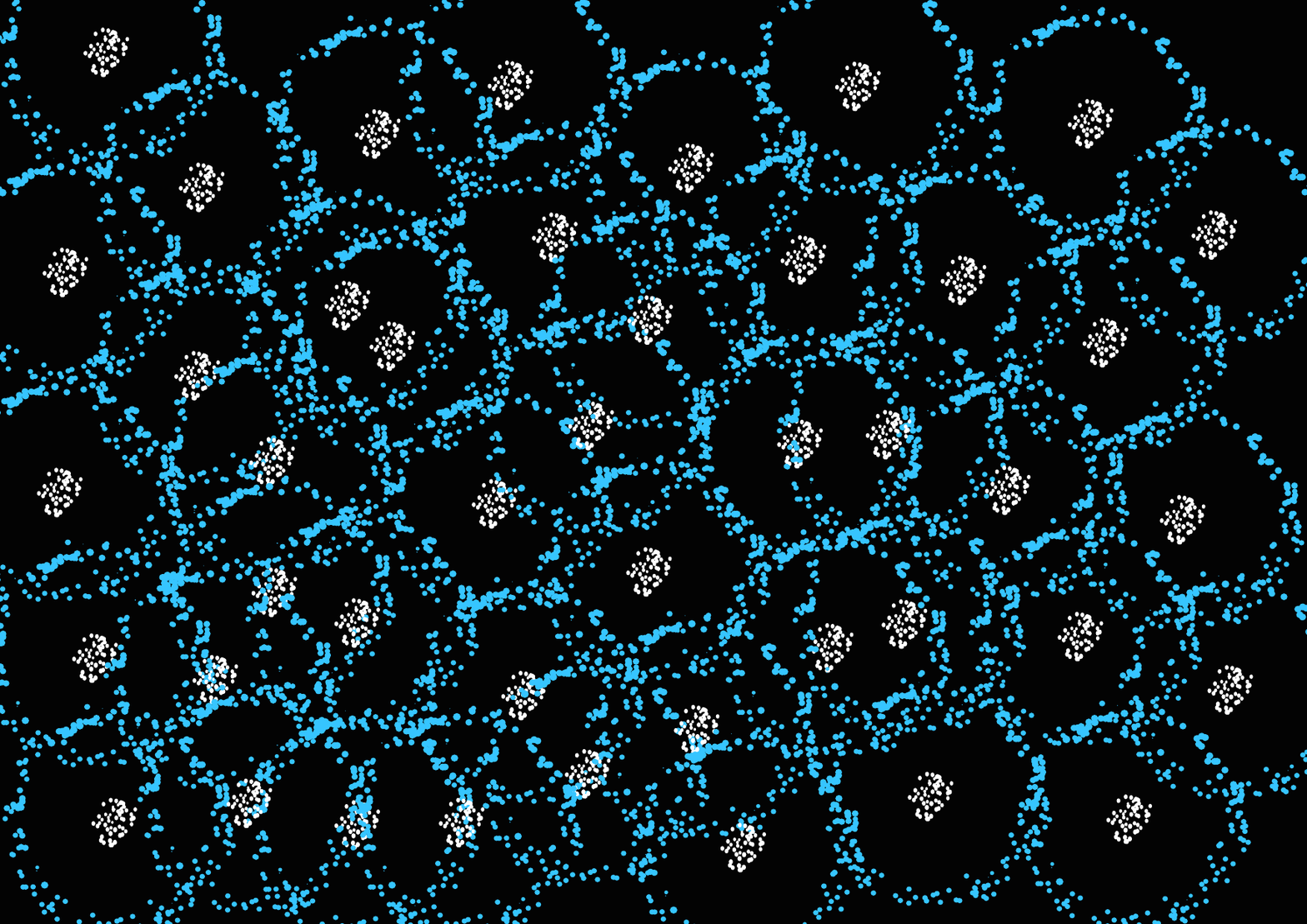}
    \caption{BAO representation from point sources. Each point can be thought of as a galaxy. The blue ones are found in the BAO feature, while the white ones are clustered due to Dark Matter after decoupling from photons.}
    \label{fig:bao_ex}
\end{figure}

In the $\Lambda$CDM model, the formation of the LSS was possible due to inhomogeneities that grew to the observed structure in the universe. Inflation was a period of accelerated expansion that could explain how quantum fluctuations stretched to form the inhomogeneities capable of growing and forming galaxies and galaxy clusters. The theory was developed by \cite{guth81} and today has its sophistication in detail in \cite{baumann2009tasi}. 

The perturbed continuity equation and the Euler equation combined lead to the result of a photon-baryon fluid, represented by the subscript $\gamma b$, we obtain
\begin{equation}\label{eq:2.52}
    \frac{1}{4} \delta_{\gamma b}'' + \frac{1}{4} \frac{R'}{1+R}\delta_{\gamma b}' + \frac{1}{4} k^{2} c_s^{2} \delta_{\gamma b} = F_k \text{ ,}
\end{equation}
\begin{equation}\label{eq:2.53}
    F_{\gamma b} \equiv -\frac{k^{2}}{3} + \frac{R'}{1+R} \Phi_b'+\Phi_b'' \text{ ,}
\end{equation}
\begin{equation}\label{eq:soundspeed}
    c_s^{2}=\dfrac{1}{3(1+R)}\text{ ,}
\end{equation}
\begin{equation}\label{eq:baryonfrac}
    R \equiv \frac{3 \rho_b}{4 \rho_{\gamma}}\text{ ,}
\end{equation}
where $F_k$, defined in Eq. (\ref{eq:2.53}), and acts as a driving term, $c_s$ is the sound speed as in Eq. (\ref{eq:soundspeed}) and $R$ is the baryon-photon ratio written in Eq. (\ref{eq:baryonfrac}).

Eq. (\ref{eq:2.53}) shows the relation between photons and baryons during a tight coupling epoch. When $c_s$ varies slowly (\ref{eq:2.53}) becomes a forced oscillator. Such coupling is sufficient for the photon-baryon fluid to oscillate as a sound wave. This oscillation is between under-dense and dense regions of photons and baryons as a spherical wave. Even though dark matter is also present at $z \approx 1000$, its interaction is essentially gravitational, thus it remains at the centre of the oscillation. The solution of (\ref{eq:2.53}) is
\begin{equation}\label{eq:2.54}
    \frac{1}{4}\delta_{\gamma b}=A_{\gamma b}(\eta)+B_{\gamma b}(\eta)cos(kr_s)+C_{\gamma b}(\eta)sin(kr_s)\text{ ,}
\end{equation}
where $r_s$ is the sound horizon. $A_{\gamma b}$, $B_{\gamma b}$, and $C_{\gamma b}$ are coefficients which vary slowly with $\eta$, in which $A_{\gamma b}$ represents the perturbation being outside the centre of the spherical wave. After decoupling, the diffusion of the photons left a pressure on the system and shells of baryons were released. The first shell released has a characteristic scale, $r_s$ and it represents the sound wave travelling from $\eta=0$ until some conformal time $\eta$,
\begin{equation}\label{eq:2.55}
    r_s=\int_0^{\eta}c_s d\eta .
\end{equation}

With $R \approx 0$, $c_s=1/\sqrt{3}$ thus, $r_s=c_s \eta$. As $\eta$ increases $R$ increases following the equation
\begin{equation}\label{eq:Reta}
    R(\eta)= \frac{3}{4} f_b (1-R_\nu)^{-1} \frac{z_{eq}}{z(\eta)}\text{ ,}
\end{equation}
where $z_{eq}$ is the redshift at  equality between matter and radiation ($1+z_{eq}=\Omega_{m0}/\Omega_{r0}$), $f_b=\rho_b/\rho_m$ and $R_\nu = \rho_\nu/\rho_r$. 

By rewriting the Friedmann equation in terms of $a_{eq}$ as written as
\begin{equation}\label{eq:2.86}
    H^{2} = \frac{1}{2} H_{eq}^{2} \left[ \left(\frac{a_{eq}}{a}\right)^{3} + \left( \frac{a_{eq}}{a} \right)^{4} \right]\text{ ,}
\end{equation}
if we integrate it in terms of conformal time the result is
\begin{equation}\label{eq:aequa}
    \frac{a(\eta)}{a_{eq}}=(2 \sqrt{2}-2)\left( \frac{\eta}{\eta_{eq}}\right) + (1-2\sqrt{2}+2) \left( \frac{\eta}{\eta_{eq}}\right)^{2}\text{ .}
\end{equation}

The use of equations (\ref{eq:aequa}), (\ref{eq:soundspeed}), (\ref{eq:baryonfrac}), and (\ref{eq:2.55}) leads to $r_s(\eta)$, written below. The sound horizon at the drag epoch corresponds to $r_s(\eta_{drag})$ when baryons were free from Compton drag. In other words, photons no longer deter the gravitational force from baryons,
\begin{equation}\label{eq:rhor}
    r_s(\eta)=\frac{3}{4}\sqrt{\left( \frac{6}{R_{eq}}\right)} \ln{\left(   \frac{\sqrt{1+R(\eta)}+\sqrt{R(\eta)+R_{eq}}}{1 + \sqrt{R_{eq}}}\right)}.
\end{equation}
If we consider the $\Lambda CDM$ model with the inflationary scenario, $z_{drag} \approx 1000$, thus $r_s(z_{drag}) \approx 150 $ Mpc. This is the BAO scale, at $z_{drag}$ photons' instabilities can no longer deter the gravitational force from baryons.

The BAO feature is imprinted in the LSS as the frozen last spherical wave-front with radius $r_s\simeq150$ Mpc. This can be observed in galaxy surveys because due to that wavefront, galaxies have a preferred scale to cluster in the shape of this feature. An example is shown in Figure~\ref{fig:bao_ex}, the dots represent one galaxy, and the blue signals are the BAO with their characteristic shape. In the centre of each BAO, there is the clustering of galaxies due to Dark Matter after decoupling from radiation.

\section{Observables and estimators}\label{sec:est}
\subsection{Two point correlation function}

In order to pursue evidence of the previous results, we need to estimate the number of objects clustered in the observable sky. For that we require an estimator which measures excess or lack of clustering.

Considering $N$ points in a volume $V$, its number density is $n=N/V$. In order to describe the distributions of points better, let us consider the infinitesimal volume $dV$, thus, $n dV$ is the average numbers in an infinitesimal volume. Taking the separation of a pair of points, $r_{ab}$  one can find the average number of pairs in the volumes $dV_a$ and $dV_b$, which is $dN_{ab}$,
\begin{equation}\label{eq:2.57}
    dN_{ab}= \langle dn_a dn_b \rangle = n^{2} dV_a dV_b[1+\xi (r_{ab})].
\end{equation}

Eq. (\ref{eq:2.57}) shows the average in a pair $ab$ with the two point correlation function $\xi(r_{ab})$, \cite{amendola2010}. If $\xi$ is zero, the distribution of particles is described by a Poisson distribution and the average of pairs is the same as the product of the average of the two volumes separately, $\langle dn_a dn_b \rangle=\langle dn_a \rangle \langle dn_b \rangle$. Therefore, the particles are uncorrelated and perfectly represent a Poissonian density field ($\delta$). On the other hand, when the particles show correlation, $\xi$ is different than zero, \cite{amendola2010}. The correlation function depends on the separation of the pairs and can be written as
\begin{equation}\label{eq:2.58}
    \xi_{ab}=\frac{dN_{ab}}{n^{2} dV_a dV_b}-1=\langle \delta(r_a) \delta(r_b) \rangle \text{ .}
\end{equation}

Eq. (\ref{eq:2.58}) means that there is an excess probability of finding a pair separated by the distance $r_{ab}$ in a distribution. The most practical way of obtaining it is setting $ndV_a=1$ and rewrite Eq. (\ref{eq:2.58}) to get
\begin{equation}\label{eq:2.59}
    \xi_{ab}=\frac{dN}{n dV}-1.
\end{equation}

Real data comes with discrete values of the position of galaxies. One must use an unbiased estimator with minimum variance, which describes the pair counts. The optimal estimator was proposed by Landy and Szalay (1993) \cite{landy1993}. The Landy-Szalay estimator estimator has the following form:
\begin{equation}\label{eq:tpcf}
\xi_{LS}= \left( \frac{N_{rand}}{N_{data}}\right)^2 \frac{DD(r_{ab})}{RR(r_{ab})} - 2 \frac{N_{rand}}{N_{data}}\frac{DR(r_{ab})}{RR(r_{ab})}+1 \text{ .}
\end{equation}

The estimator is made of the number of pairs in the real catalog, $DD$, the random one, $RR$, and the galaxy-random pair $DR$. $RR$ is a pair of an artificial galaxy catalog that forms a Poisson sampling, it is distributed in the same boundaries as the real catalog.

$N_{rand}$ is the number of galaxies in the random catalog, while $N_{data}$ is the number of galaxies in the real catalog, they are inserted in order to normalize the pair counts when the catalogs have different size. Moreover, the lack of distribution of real data could lead to a poor representation of the BAO echo, it then is advantageous to have $N_{rand}>N_{data}$.

\subsection{Matter Power Spectrum}\label{sec:power}

It is possible to obtain the matter distribution of the universe by either calculating the matter power spectrum or the correlation function. As $\delta$ has zero average, it is convenient to carry statistics of quadratic functions. The quadratic functions of the perturbations($\delta$) are called power spectrum $P(k$), which are simply the variances($\langle \delta(\vec{k})\delta(\vec{k}')\rangle$) of the Fourier modes.

The variance can be written as follows:
\begin{equation}\label{eq:defpk}
    \langle \delta(\vec{k})\delta(\vec{k}')\rangle = (2 \pi)^3 \delta^{(3)}(\vec{k}-\vec{k'}) P(k) \text{ .}
\end{equation}

The reader should note that $\delta(\vec{k})$ should lead to $P(\vec{k})$; however, when considering isotropy these quantities should depend on the module of the vectors $\vec{k}$ and $\vec{r}$. Furthermore, the power spectrum is the Fourier transform of the correlation function, which is
\begin{equation}\label{eq:pk0}
    P(k) =  \int \xi(\vec{r}) e^{-i\vec{r} \cdot \vec{k}} dV = \int \xi(r) e^{-ir \cdot k} dV   \text{ .}
\end{equation}

The power spectrum and $\xi$ are equivalent to each other in theory. However, in terms of observational data, they are not the same, since the estimators are susceptible to the errors of the observations and we do not have access to all scales. The two-point correlation function measures the excess probability of finding a pair of galaxies separated by some distance, while the power spectrum assesses the contributions of $\delta$ on different scales.

In the physical world, the BAO is a 3D feature, but when the power spectrum monopole is used, one can extract the 1D information of the BAO. The isotropic, one-dimensional version of the scale, $D_V$, was described by \cite{eisenstein2005detection}. Depending on where to look at the literature, $D_V$ is called the dilation scale or 1D BAO scale or isotropic BAO distance, and is given by
\begin{equation}\label{eq:dilation}
    D_V(z) = \left[(1+z)^2 D_A(z)^{2} \frac{cz}{H(z)}\right]^{1/3},
\end{equation}
where $z$ is the redshift, $D_A(z)$ is the angular diameter distance, and $H(z)$ is the Hubble parameter. $D_V(z)$ can be thought of as a geometric mean of the distances in independent directions in 3D space, two transverse and one along line-of-sight. 

\subsection{Angular quantities}

Instead of analysing the BAO feature as a three-dimensional probe, one can observe its transverse to the line-of-sight scale. The relation between an angular power spectrum and the matter power spectrum is written as follows:
\begin{equation}\label{eq:cell_ini}
    C_\ell^{A,B}=\frac{2}{\pi}\int \mathrm{d} k k^2P(k)\Delta_\ell(k) \Delta_\ell(k),
\end{equation}
where $\Delta_\ell(k)$ is
\begin{equation}
    \Delta_\ell(k)=\langle\delta\delta^* \rangle.
\end{equation}
These terms depend on the cosmological parameters of interest.

\begin{equation}
    w(\theta)=\langle \delta(\hat{\mathbf{n}})\delta(\hat{\mathbf{n}}')\rangle
\end{equation}

here $\hat{n} \cdot \hat{n}'=\theta$, the angular separation between a pair of objects. $\delta(\hat{\mathbf{n}}),\delta(\hat{\mathbf{n}}')$ are the density fluctuations in of a particular pixel $\hat{\mathbf{n}}$. Using properties of spherical harmonics,
\begin{equation}
    \delta(\hat{n})=\sum_{\ell m} a_{\ell m} Y_{\ell m}(\Theta,\Phi).
\end{equation}
$a_{\ell m}$ are multipole moments. Because the angular power spectrum is $C_\ell = \sum_{m=-\ell}^\ell |a_{\ell m}|^2/(2\ell +1)$, its Fourier transform is written as
\begin{equation}
    w_{ij}(\theta) = \sum_\ell \frac{(2\ell+1)}{4 \pi } C_\ell^{ij} P_\ell(\cos{\theta}),
\end{equation}

\subsection{Galaxy bias}
Galaxies are a good representation of baryonic matter, but we are ultimately interested in the CDM distribution which is not directly observable via electromagnetic signals. A way of compensating for the lack of information is the introduction of a quantity to estimate the difference between the two types of matter. The galaxy bias $b$ was introduced to estimate the distribution of galaxies in terms of total matter. The concept was introduced by Kaiser (1984) \cite{kaiser1984spatial}. Once galaxies and CDM are not the same, the matter power spectrum of galaxies needs this estimation.

The power spectrum of galaxies can be written as
\begin{equation}
    P_g =  b^{2} P_{\delta_m}\text{ ,}
\end{equation}
$b$ is the galaxy bias and it implies that the galaxy density is not exactly equal to the matter density since the galaxies tend to form in regions of higher CDM density. The same can be implied in the two-point correlation function $\xi$:
\begin{equation}
    b^2=\frac{\xi_{g}}{\xi_m}\text{ .}
\end{equation}

The galaxy bias is a statistical relation destined to indicate whether the mean galaxy density($\bar{n}_g(\textbf{r})$) is linearly proportional to the matter-energy density($\rho_m(\textbf{r})$). If $b<1$, galaxies are less clustered than dark matter. The equation with such a relation is written as
\begin{equation}
    \delta_g(\textbf{r}) \equiv \frac{n_g(\textbf{r})}{\bar{n}_g}-1 = b\delta(\textbf{r})= b \left( \frac{\rho_m(\textbf{r})}{\bar{\rho}_m}- 1 \right)\text{ ,}
\end{equation}
where $n_g(r)$ is the galaxy number density within a radius $r$, $\bar{n}_g$ is the mean number density.

There are more sophisticated models in the literature, including a possible variation with redshift. In this work, we will consider the simplest description, with a constant bias.

\section{Mock catalogs}\label{sec:mocks}

Surveys have their own characteristics, either because of their precision or because of the type of objects used. Each analysis carries the necessity to validate their methods. To do so, we need a model, like any physics phenomenon. In the case of surveys, the perfect model would be the impossible task of simulating the formation of the universe and the LSS. The closest we can get to such a task is through N-body simulations.

The universe's four fundamental forces are not well understood yet, even though there is strong evidence for them, with solid experiments to prove them. However, their role in each phase of the formation of the LSS is too complex to be carried in a numerical simulation. N-body simulations try to find the closest to a simulated universe by accounting mostly for dark matter's contribution to the LSS.

Assuming CDM as the main species in the universe, it is simple to reduce the Friedman equations to a final equation of motion \parencite{bagla2005cosmological}. Modern simulations also take advantage of observations made in the last decades. The Millennium Simulation  \parencite{springel2005simulations} is the main example of the combination of increasing knowledge with models, simulating $2,160^3$ dark matter particles from redshift $z=127$ to $z=0$.

To validate the algorithms that detect BAO in surveys, we require a good representation of a survey from simulations. The N-body simulations do not have galaxy bias due to their dark matter construction, so it is important to construct mock catalogues from the N-body simulations using functions that mimic the properties of real surveys. 

An important "biasing" effect is the position of the observer. Depending on the location of the observer, whether in a rich galaxy cluster or not. This was described at 
\cite{cole1998mock} for the 2dF mock catalog.

The mocks for the SDSS-III generation were built from the MultiDark Simulations \parencite{klypin2016multidark}. The MultiDark Patchy Mocks \parencite{kitaura2016clustering} presents an accurate galaxy bias, growth of structures, and RSDs which resulted in accurate BAO reconstruction. 

For the SDSS-IV generation, the SDSS Collaboration chose less computationally expensive mock sets. The LRG, QSO, and ELG mocks are made from the effective Zel’Dovich approximation mock (EZmock) algorithm based on the Z'eldovich approximation \parencite{zel1970gravitational} described at \cite{zhao2021completed}.

\section{Spectroscopic and photometric surveys}

The spectroscopic redshift ($z$) is based on the actual frequency shift in the lines of the spectrum emitted from any source. They offer precise measurements of the redshift, providing crucial results for cosmological analysis. The light passes through slits that ensure a limit to the number of spectra detected. Spectrographs split light from sources into many frequencies and then send them to a photo-detector. Splitting the light and the reflection from the mirrors of the telescope makes faint objects difficult to observe, so they usually need a longer exposure time to be measured. 

The first detection of BAO was made by \cite{eisenstein2005detection1}. The data comprised the first sample of Luminous Red Galaxies (LRG) from the Sloan Digital Sky Server (SDSS) spectroscopic survey. The evidence was collected through the two-point correlation function, and the result indicated the position of the BAO peak at $z = 0.35$, the effective redshift of the sample. At the same time, the 2dF Galaxy Redshift Survey (2dFGRS) \cite{cole20052df} found the BAO signal using the 3D power spectrum, being the first results with matter content in the universe using this type of probe.

Since then, the SDSS cumulative mission has observed more objects. The latest results were from SDSS-IV Data Release 17 (DR17) \cite{accetta2022seventeenth}. The Extended Baryon Oscillation Spectroscopic Survey (eBOSS) sample includes LRG, emission line galaxies (ELG), and quasi-stellar objects (QSO). The DR16 \cite{wang2020clustering} results of LRG and ELG for $z=0.77$ found an angular diameter distance of $D_A/r_s = 18.85 \pm 0.38$.

The WiggleZ Dark Energy Survey is another spectroscopic survey of ELG \cite{drinkwater2010wigglez}. \cite{blake2012wigglez} obtained the $D_A = 1205 \pm 114, \text{ } 1380 \pm 95, \text{ } 1534 \pm 107 $ Mpc for $z= 0.44,\text{ } 0.6 \text{ and } 0.73$. 

Due to the nature of the BAO signal, it is important to have as many galaxies as possible to constrain the feature. Photometric surveys have the advantage of collecting large samples of old galaxies. 

Photometric observations are collected from filters sensitive to certain wavelengths. Each object observed has a flux/magnitude value for each filter. This is called the photometry of an object. The information is then used to estimate the redshift of each object called the photometric redshift (photo-z). 

However, photometric redshifts (photo-z) are not as precise as spectroscopic measurements due to the reliability of integrated fluxes from the filters and a representative spectroscopic sample \cite{salvato2019many}, used to train the photo-z algorithms. This leads to redshift uncertainties. Usually, a survey's accuracy is $\sigma = \sigma_{survey} (1+z)$, the Dark Energy Survey (DES) uncertainty is about $\sigma \sim 0.03 (1+z)$ \cite{crocce2019dark}. 
\begin{figure}
    \centering
    \includegraphics[width=.8\textwidth]{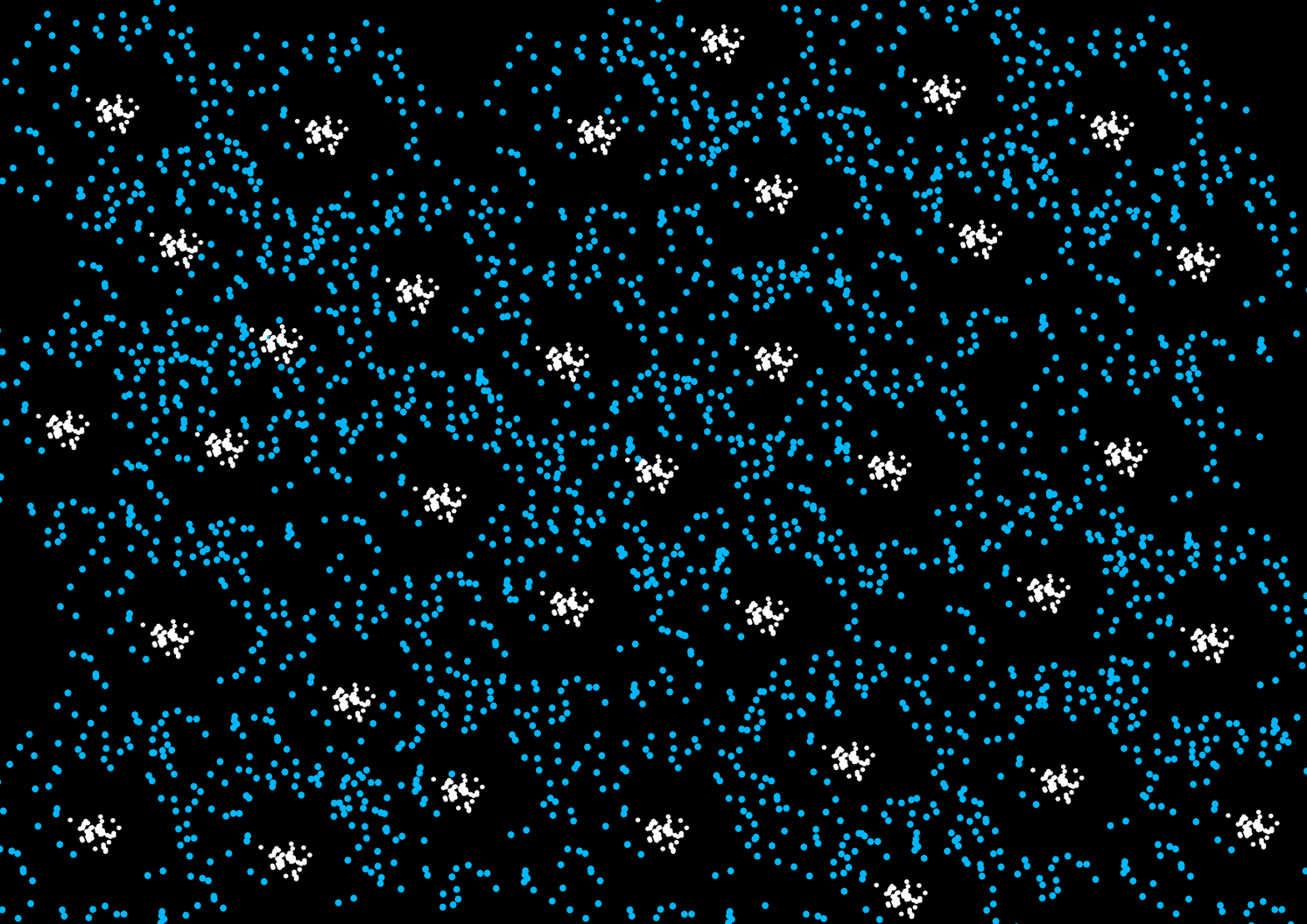}
    \caption{BAO representation from point sources measured by photometric redshift. Each point can be thought of as a galaxy. The blue ones are found in the BAO feature, while the white ones are clustered due to Dark Matter after decoupling from photons.}
    \label{fig:pzbao}
\end{figure}

\begin{figure}
    \centering
    \includegraphics[width=.8\textwidth]{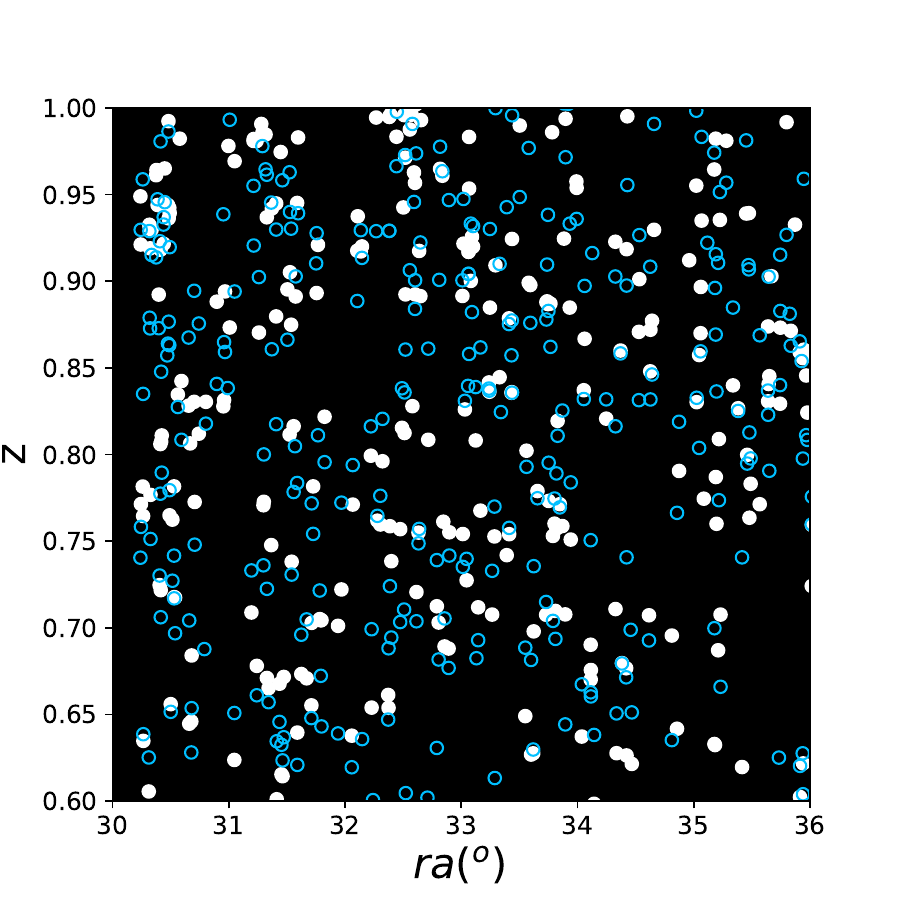}
    \caption{Example of photometric redshift estimation. In white, there are sources from spectroscopic surveys and in blue, the same sources from the Dark Energy Survey (DES) Year 3 BAO sample \cite{DESY3}.}
    \label{fig:pz_example}
\end{figure}

High redshift uncertainties weaken the BAO signal in the line of sight. The BAO feature is smeared by photometric redshift, this is illustrated in Figure~\ref{fig:pzbao}. The exact positions of the galaxies are now fuzzy because the photo-zs could be underestimated or overestimated. An example of this problem is shown in Figure~\ref{fig:pz_example}. in white there are sources from spectroscopic surveys and in blue, the same sources from the Dark Energy Survey (DES) Year 3 BAO sample \cite{DESY3}. The spectroscopic information was matched with the VIMOS Public Extragalactic Redshift Survey Data Release 2 (PDR-2) \cite{scodeggio2018vimos} both W1 and W4 equatorial fields, DEEP2  Galaxy Redshift Survey \cite{newman2013deep2}, VIMOS VLT Deep Survey (VVDS) \cite{le2005vimos}, and the SDSS eBOSS LRG pCMASS \cite{wang2020clustering}. We see that the galaxies do not match the photo-z to their respective precise spec-z. Thankfully, the BAO is a three-dimensional feature, the transverse part is not impacted by the redshift uncertainty. The usual solution is to use angular clustering to find the BAO scale either the two-point angular correlation function or the angular power spectrum. 

\section{Constraining the BAO in 3D}\label{sec:method}
\subsection{Data and mocks}

The data used for this work are the DR12 from the SDSS, which includes an LSS catalog of BOSS spectroscopic observations. These observations were obtained through two target selections: the LOWZ sample targets luminous red galaxies with $z<0.4$ and CMASS targets massive galaxies with $0.4<z<0.7$, \parencite{reid2016sdss}. We used both LOWZ and CMASS v5 \footnotetext{BOSS data:  \hyperlink{https://data.sdss.org/sas/dr12/boss/lss/}{https://data.sdss.org/sas/dr12/boss/lss/}}, comprising a total number of galaxies observed is 993,228 with $0.3<z<0.65$.

In order to calculate the power spectrum covariance matrix, we need simulations of the BOSS dataset. Each simulation, called mock catalog, represents a simulated universe that is distributed in the sky as the real data, assuming a fiducial cosmological model and with the same redshift distribution. Here, the set used combines CMASS and LOWZ. The data is available in the FITS format which is what is required for $\texttt{nbodykit}$'s environment, while the mocks are in the CSV format, which needs to be converted before use.

We used 500 mocks from \cite{kitaura2016clustering,rodriguez2016clustering}, the
MultiDark-Patchy mocks\footnotetext{MultiDark-Patchy mocks:  \hyperlink{http://www.skiesanduniverses.org/page/page-3/page-15/}{http://www.skiesanduniverses.org/page/page-3/page-15/}}. The same random set for both Northern (NGC) and Southern (SGC) Galactic Caps was used for all simulations, 50 times the size of the mocks. The catalogs were generated with the following cosmological parameters consistent to \cite{ade2016planck} cosmology. 

\subsection{Power Spectrum}

One needs to calculate a comoving distance $r$, since the data set is measured in the redshift space. The distance $r$ is only obtained assuming a fiducial cosmological model, for which we chose a flat $\Lambda$CDM model with the Planck 18 results \cite{collaboration2018planck}. The same model was applied to the mocks, which is summarized in the table \ref{tab:fiducial} together with other important quantities, such as the mean temperature of the CMB ($T_{CMB}$), the spectral index $n_s$ and the number of ultra-relativistic neutrinos($N_{ur}$). 

We show the 3D distribution of the galaxies in Mpc, including the redshift information in a colour scale, the larger the redshift, the lighter the blue and the higher the distance from the centre, which is the observer. The smaller portion is the SGC and the larger portion is the NGC. Because we are using a slice of DR12, the galaxies are concentrated in two slices of spherical cones. The entire data set would display two cones in each galactic cap.

To obtain the matter power spectrum of a galaxy catalog, we perform the calculation with $\texttt{nbodykit}$'s $\texttt{FKPCatalog}$ class to convert the catalog into an FKP catalog which is the same catalog in Fourier space according to a fiducial model. Then, the $\texttt{ConvolvedFFTPower}$\footnotetext{The algorithm is described in nbodykit's website: \url{https://nbodykit.readthedocs.io/en/latest/results/algorithms/survey-power.html?highlight=fft\%20convolved\#the-algorithm}.} class uses the method to estimate the power spectrum of \cite{feldman1993power}. The procedure involves computing the Fourier transform of the weighted galaxy fluctuation field $F(\vec{k})$. From $F(\vec{k})$, the estimator obtains the power spectrum from the data $\hat{P}(\vec{k})$ by subtracting the contribution from the shot noise ($P_{shot}$) which is described in \cite{feldman1993power}.

The power spectrum is calculated using a $512^3$ grid with the following characteristics, shown in Table \ref{tab:sgc_ngc_pk} (\cite{zhao2016clustering} used $1024^3$ cubic cells for the same volume). The size of the boxes, in comoving coordinates, of the NGC is $1596$ Mpc $h^{-1}\times3012$ Mpc $h^{-1} \times1666$ Mpc $h^{-1}$, and the SGC is $1082$ Mpc $h^{-1} \times 2328$ Mpc $h^{-1} \times 1308$ Mpc $h^{-1}$. Furthermore, the wave number ranges from $k=0.01 h$ Mpc$^{-1}$ to $k=0.3h$ Mpc$^{-1}$.

\begin{table}[ht]
\centering
\caption{\label{tab:fiducial}Fiducial model used to calculate comoving distances.}
\begin{tabular}{|c|}
\hline
$\Omega_{\Lambda}= 1- \Omega_{CDM}- \Omega_{b}$ \\
$\Omega_{CDM} = 0.179$ \\ 
$\Omega_{b} = 0.033$   \\ 
$h=0.6727$              \\ 
$n_s = 0.9649$          \\ 
$T_{CMB}=  2.7255$ K \\ 
$N_{ur} = 2.0328$       \\
gauge = synchronous   \\ 
\hline
\end{tabular}

\end{table}
The power spectrum obtained for the SGC and NGC has the following attributes: the wave number separation ($dk$), the Poisson shot noise, the number of points, the number of random points, the size of each box and caps, for which we consider only the monopole. The values are summarised in table \ref{tab:sgc_ngc_pk}. 

 \begin{table}[ht]
 \centering
    \caption{\label{tab:sgc_ngc_pk} Attributes obtained after calculating the power spectrum of each galactic cap.}
    \begin{tabular}{|l|l|l|l|}
    \hline
     &NGC BOSS& SGC BOSS & eBOSS\\
     \hline
    $dk$(h Mpc$^{-1}$) & 0.01 & 0.01 & 0.01 \\
    Poisson shot noise(h/Mpc$^{3}$) &4240.56 & 4201.62 & 16,070.32\\
    Number of points &720,113 & 273,115 &174,816\\
    Number of randoms &37,115,901& 13,647,332&8,914,172\\
    $N_{mesh}$ &$512^3$ & $512^3$ & $512^3$ \\
    Size of box(Mpc/h) & 1596, 3012, 1666 & 1082, 2328, 1308 & 4373, 4030, 2308\\
    Poles & $P_0(k)$    & $P_0(k)$ & $P_0(k)$ \\
    \hline
    \end{tabular}
    \end{table}

It is important to distinguish the Poisson shot noise from shot noise ($P_{shot}$). The poisson shot noise is the ratio between the volume of the simulation box and the number of objects, assuming a perfectly Poissonian distribution. The $P_{shot}$ is the actual result of this error obtained from a given sample. It requires a weighted estimation of the shot noise, considering the volume of the box and the individual galaxies' weights ($w_{tot,i}$):

\begin{equation}
    P_{shot} = V \frac{\sum_i w_{tot,i}^2}{(\sum_i w_{tot,i})^2}.
\end{equation}

The same routine was applied to mocks whose attributes will not be explicitly shown, since there are 500 mocks from \cite{kitaura2016clustering} and \cite{rodriguez2016clustering}. Each has its own attributes similar to the data set.

\subsection{The BAO feature with the matter power spectrum monopole}\label{sec:bao_feature}

It is common to use a polynomial fit for the wiggles of the power spectrum. To do so, it is useful to write two functions, a smoothing model representing a power spectrum without the BAO and the other including the BAO in the power spectrum using the same fiducial model used to obtain the comoving distance. This method is the same as the one used in \cite{anderson2014clustering}.

The smooth model is defined as
\begin{equation}\label{eq:pk_smooth}
    P_{sm}(k)=b^2 P_{noBAO,lin}(k)+ A_1 k+A_2+\frac{A_3}{k}+\frac{A_4}{k^2}+\frac{A_5}{k^3} 
\end{equation}
where $A_1,A_2,A_3,A_4,A_5$ are the polynomial coefficients that do not have an explicit physical meaning, and the parameter $b$ is the constant large-scale bias. The polynomial parameters are used to fit small-scale effects \cite{anderson2014clustering}. The linear power spectrum without BAO $P_{noBAO,lin}(k)$ is based on the fitting formulae from \cite{eisenstein1998baryonic}, as implemented in $\texttt{nbodykit}$.  $b$ is the galaxy bias, the parameter is ideally close to one, which we kept as a free parameter in our study.

The model that contains the oscillations is described as
\begin{equation}\label{eq:pkfit}
    P_{fit}(k) = P_{sm}(k) [1+(Osc_{lin}(k/\alpha)-1)e^{-0.5k^2 \Sigma_{nl}^2}],
\end{equation}
where $Osc_{lin}$ is the purely oscillating part of the power spectrum , also calculated using  \cite{eisenstein1998baryonic} fitting formulae in $\texttt{nbodykit}$. The additional parameter $\Sigma_{nl}$ is related to the damping scale of the oscillations, while $\alpha$ is the scale dilation parameter given by
   \begin{equation}\label{eq:alpha}
        \alpha = \frac{D_V(z)}{D_{V,fid}(z)}\text{,}
    \end{equation}
where $D_{V}$ and $D_{V,fid}$ are the dilation scales of Eq. (\ref{eq:dilation}), both from the data set and the fiducial model. 

We fit the model for $\alpha$, $\Sigma_{nl}$, bias, and the five polynomial coefficients using flat priors. In particular, we chose $0.8<\alpha<1.2$, as in \cite{anderson2014clustering}, and $0 h^{-1} Mpc<\Sigma_{nl}<20h^{-1} $ Mpc, given the CMASS and LOWZ samples characteristics as described by \cite{anderson2014clustering}.

\subsection{Covariance matrix}

The mock catalogs were used to compute the covariance matrix, which is used later for fitting a model to the data. The mocks are a means of evaluating the model of the power spectrum and a way to estimate its error bars. Given a set of survey realizations, how much of the model we fit deviates from the mean. They represent a statistical distribution of universes of the observed region of our data set with the same fiducial model. Thus, a better statistical representation of the galaxy survey requires a larger number of mocks.

The covariance matrix of the $N_{mock}=500$ mocks is
\begin{equation}\label{eq:cov_matrix}
    C_{ij} = \frac{1}{N_{mock}-1} \sum_{p=1}^{N_{mock}} [ P_p(k_i) - \bar{P}(k_i)] \times [P_{p}(k_j)-\bar{P}(k_j)] 
\end{equation}
where $p$ to represent each mock the term $\bar{P}(k_i)$ is the average value of the power spectrum of all mocks used and it is written as
\begin{equation}
    \bar{P}(k_i) = \frac{1}{N_{mock}} \sum_{p=1}^{N_{mock}} P_p(k_i)
\end{equation}

\section{Results}\label{sec:results}
\subsection{Best-fit}
\begin{figure}
\centering
\includegraphics[width=.8\textwidth]{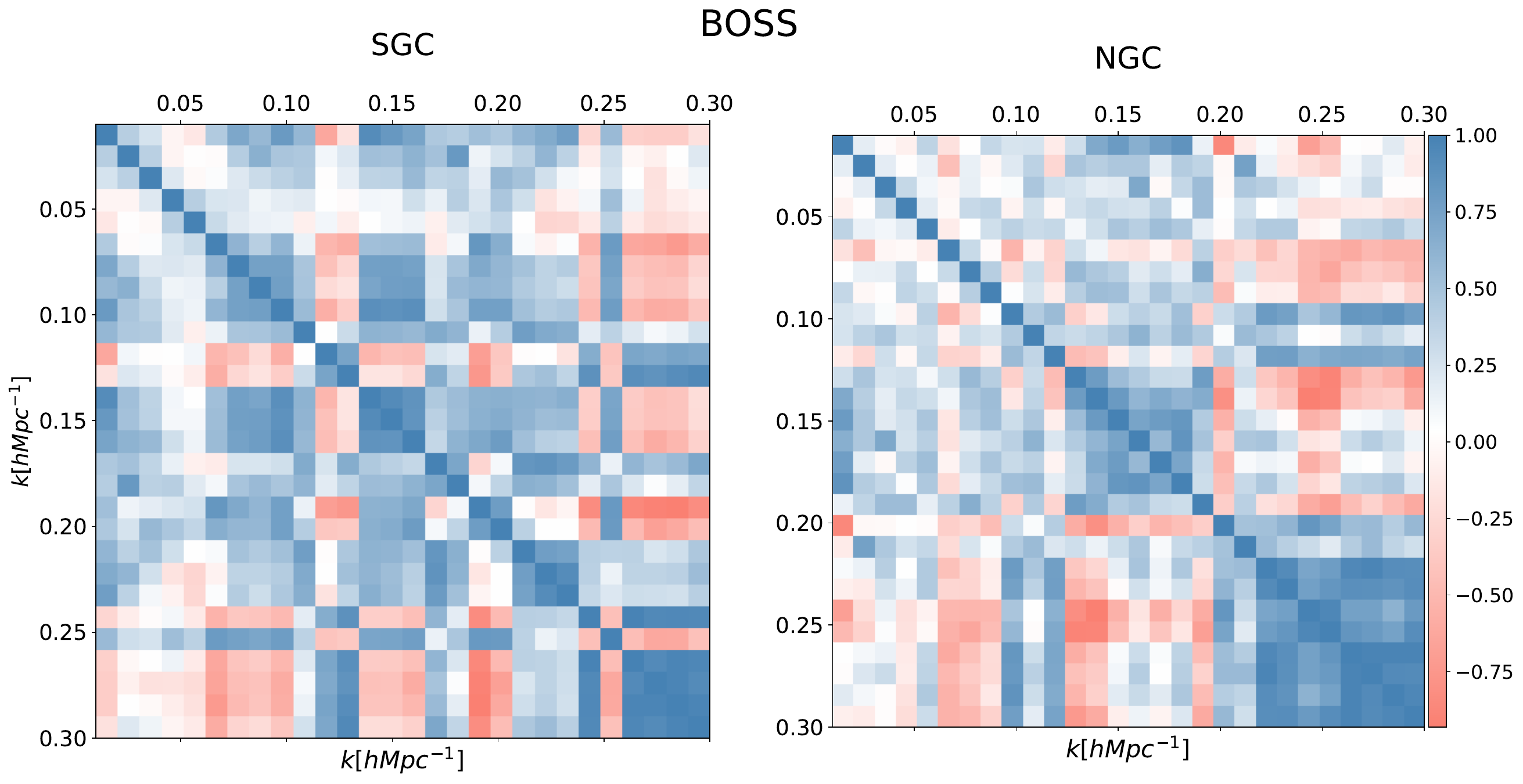}
\caption{The correlation matrix relative to the covariance matrix of the power spectrum of 500 mocks. Correlated values approach the blue colour, uncorrelated values are in yellow, and negative correlation is in red.}
\label{fig:cov_matrix0}
\end{figure}

After implementing the methodology of \cite{anderson2014clustering}, the BAO feature can be obtained. Firstly, the estimated power spectrum obtained from the DR12 data is shown in Figure \ref{fig:cov_matrix0}, the covariance matrix of SGC and NGC is represented in the left and right panels of the BOSS set. The correlated values of the wave number are represented in blue, the uncorrelated values are represented in yellow, and the negative correlation is depicted in red. The difference between the results for each galactic cap may be explained by sample size. The SGC is noisy due to the smaller number of objects. The possible reason for so many highly correlated/anticorrelated values outside the diagonal is that we performed a computation with only 500 mocks as used in \cite{marra2019}. The SDSS Collaboration used 2048 mocks to calculate their covariance matrices in \cite{zhao2016clustering} which is computationally expansive. It is also evident that the correlation is on smaller scales in the bottom right of both panels; this shows that a smaller sample of catalog data will display small-scale effects. 

\begin{figure}
\centering
\includegraphics[width=.5\textwidth]{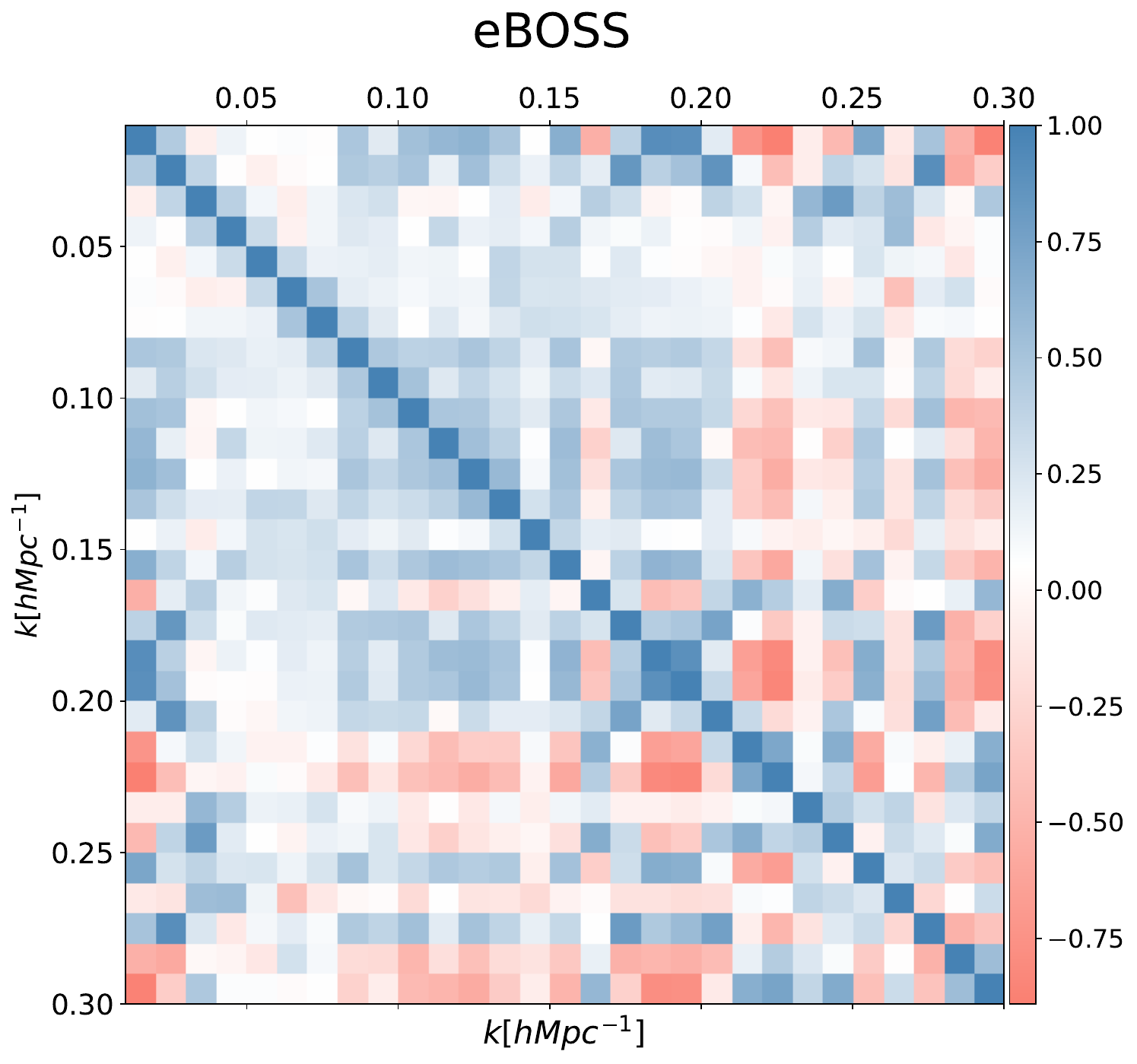}
\caption{The correlation matrix relative to the covariance matrix of the power spectrum of 500 mocks. Correlated values approach the blue colour, uncorrelated values are in yellow, and negative correlation is in red.}
\label{fig:cov_matrix2}
\end{figure}

The same visual representation was applied to the eBOSS set in Figure~\ref{fig:cov_matrix2}, this time, there is much less correlation in off-diagonal elements, this is because the redshift range is higher and less susceptible to non-linear effects.

\begin{figure}[ht]
\centering
\includegraphics[width=.8\textwidth]{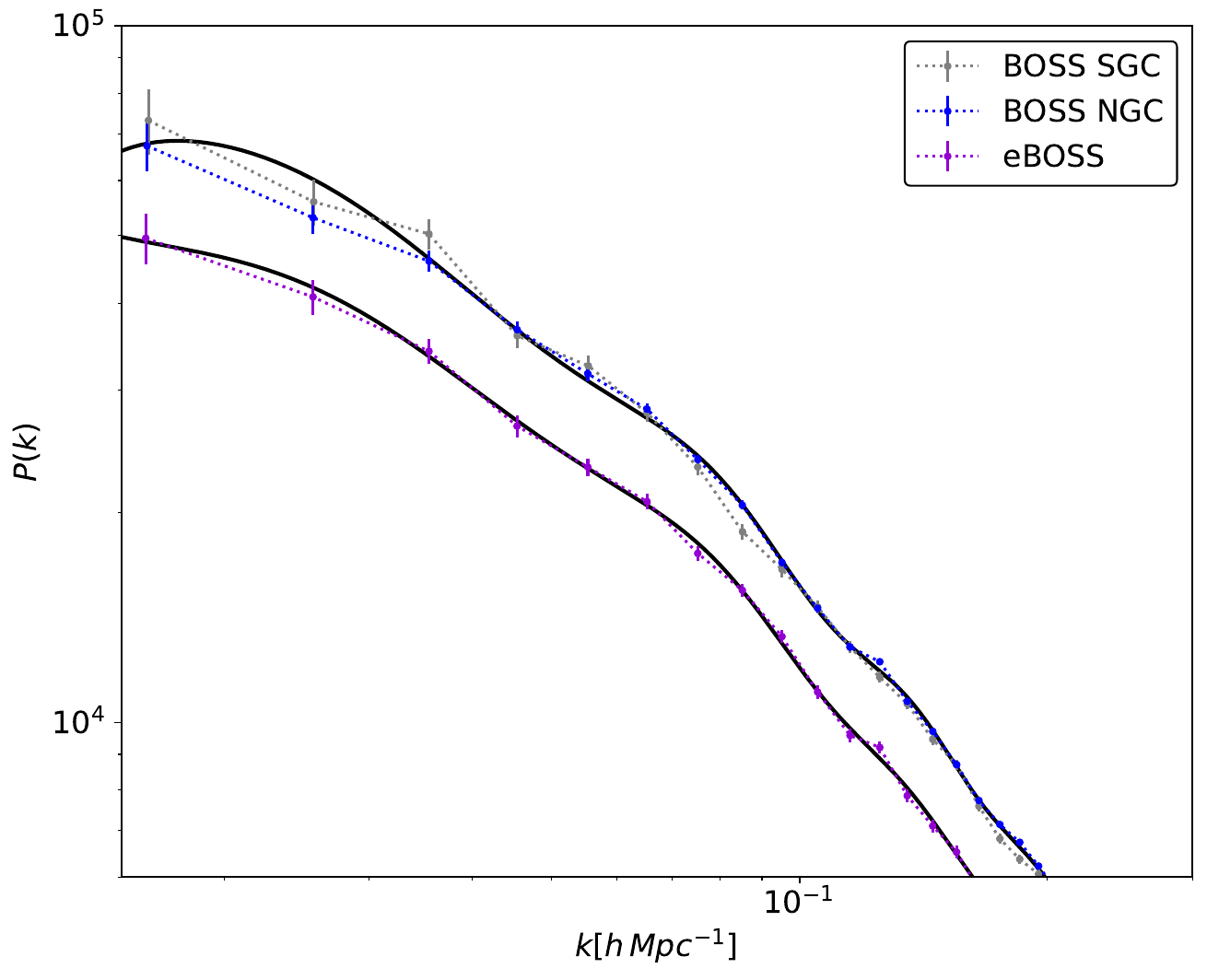}
\caption{Power spectra with the two samples. \textit{Blue dots}: BOSS NGC. \textit{Gray dots}: BOSS SGC. \textit{Pink}: eBOSS. \textit{Black}: fitted model.}
\label{fig:pk}
\end{figure}

To obtain the best fit, we use the Monte Carlo Markov Chain (MCMC) method, as implemented in the python package $\texttt{emcee}$\cite{foreman2013emcee}. Equations (\ref{eq:pk_smooth}) and (\ref{eq:pkfit}) together depend on 8 parameters, 6 nuisance ones ($b,A_1,A_2,A_3,A_4,A_5$) and the remaining with flat priors ($\alpha$ and $\Sigma_{NL}$). For the eBOSS dataset, we dropped the parameter $A_5$ because this dataset has a higher redshift range, so it requires fewer nonlinear corrections. The resulting fitted power spectrum is shown in Figure~\ref{fig:pk}. We know from the definition in Eq.~(\ref{eq:defpk}) that $P(k)\propto 1-\frac{1}{(1+z)^6}$, so the higher effective redshift has the smaller $P(k)$, this is evident in Figure~\ref{fig:pk}.

The MCMC method used here had 2,000 walkers in the ensemble; 10,000 steps in which the initial state comes from the minimisation of the chi-squared function of the model using the minimiser $\texttt{iminuit}$\cite{pearcescikit}. We fitted power spectra for the combined galactic caps. The result for the whole sample was the best fit with the chi-squared $\chi^2\simeq 93$, while the number of degrees of freedom is $dof=50$ for the combined caps, while NGC and SGC got $\chi^2/dof \simeq 33/21$ and $\chi^2/dof \simeq 35/21$, respectively. eBOSS had a $\chi^2=21.59/35$, and visually, the fit seems superior to BOSS's. The same wave-number spacing was applied.

The $\chi^2$ is not as good as found in the tomographic measurements of the BOSS DR12 in \cite{zhao2016clustering}. The reason for this is the lack of points, once our wave-number spacing is large $\Delta k = 0.01$ h Mpc$^{-1}$. Furthermore, we used fewer points to form a mesh grid compared to the SDSS team, they used $N_{mesh}=1024^3$. Another important difference is the number of mocks used in this study, 500, while \cite{zhao2016clustering} used all the 2048 mocks from  \cite{kitaura2016clustering} and \cite{rodriguez2016clustering}.

\begin{figure}[ht]
\centering
\includegraphics[width=.8\textwidth]{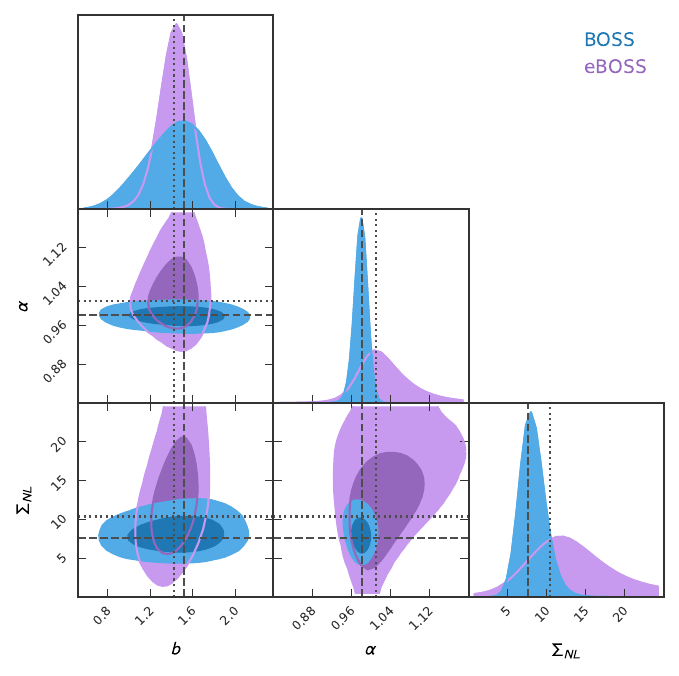}
\caption{Triangular plot for the parameters $\alpha$, $\Sigma_{NL}$ and $b$. \textit{Blue}: BOSS constraints. \textit{Purple}: eBOSS constraints.}
\label{fig:triangle}
\end{figure}
\subsection{Physical results}
Figure \ref{fig:triangle} shows the results of the likelihood distribution of the physical parameters $b$, $\alpha$, and $\Sigma_{NL}$ when using both galactic caps. Given the MCMC posterior distributions, we estimate the errors of each parameter. The diagonal plots show the one-dimensional distributions of each parameter in the 68\% confidence level. The off-diagonal plots are the correlations between the parameters, the internal contours are the $1 \sigma$ interval of the two-dimensional likelihood distribution, the external contours are the $2 \sigma$ ones.

The parameter $\alpha$, defined in Eq.~\ref{eq:alpha}, represents how much the results deviate from a fiducial model where the ideal result is $\alpha\sim1$. We chose to compare this parameter with other publications instead of computing $D_V$ which is common in the literature, the dilation scale requires a fiducial model of choice which is not as independent as it should be.

$\Sigma_{NL}$ shows the strength of the wiggles dampening and lastly, $b$ shows how much of the results are close to the linear regime. In blue, we see that the two samples agree in 1$\sigma$ for the three physical parameters, but they present a few differences. BOSS has a larger sample, so the precision in $\alpha$ and $\Sigma_{NL}$ is higher than eBOSS. However, for smaller redshifts, the bias precision is smaller than eBOSS's.

One interesting aspect is that we had to include a fiducial model to compute comoving distances and obtain $k$. Moreover, the mocks themselves had to be constructed based on a fiducial model. This could include systematics in the resulting BAO feature, but in most works the main argument is that the survey statistics ensure robustness to the estimation. A recent test by \cite{perez2024fiducial} confirmed that a Dark Energy Spectroscopic Instrument (DESI) like survey has enough statistics to overcome this systematics.

In the literature, there are attempts to constrain the BAO feature without any fiducial model using the correlation function as a function of redshift/redshift separation like in \cite{sanchez2011tracing}, \cite{sanchez2013precise},  \cite{marra2019}, \cite{menote2022baryon}, \cite{ferreira2024angular}.  \cite{marra2019} and \cite{ferreira2024angular} studied the possibility of getting the parameter results without using a mock catalog, but instead a covariance matrix from the data set. In terms of the main parameter, $\alpha$, \cite{carvalho2016baryon} proposed using \cite{sanchez2011tracing} polynomial to fit the function and then correcting to the BAO angular feature using $\alpha$ as a deviation between binning and not binning the survey. Later, \cite{ferreira2024angular} made use of thin bins to make a similar correction based on statistical information from a $\alpha$ distribution. Attempts to avoid too much fiducial information are still in progress among the community.

\begin{figure}[ht]
\centering
\includegraphics[width=.8\textwidth]{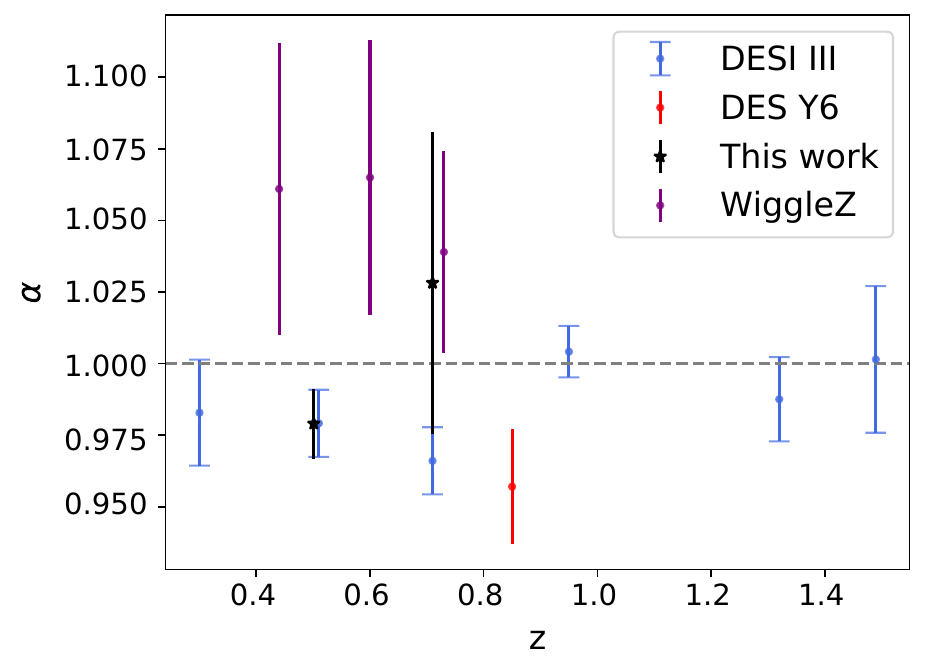}
\caption{$\alpha$ w.r.t. $z$ from DESI (blue), DES (red), WiggleZ (purple), and this work (black).}
\label{fig:alpha}
\end{figure}

We compared our results with a collection of $\alpha$ from many surveys, shown in Figure~\ref{fig:alpha}. The blue point belongs to the latest spectroscopic BAO analysis the Dark Energy Spectroscopic Instrument (DESI) III \cite{adame2024desi} which includes the BGS, LRG1, LRG2, LRG3, ELG1, LRG3+ELG1, QSO. The Dark Energy Survey(DES) \cite{abbott2024dark} final result is shown in red, it has larger error bars than DESI because it is a photometric sample. Another photometric result comes from the WiggleZ Dark Energy Survey \cite{kazin2014wigglez} with even larger error bars than DES because it has fewer objects.

Our results are the stars in black. The smaller $z$ star is the BOSS dataset $\alpha$, it has smaller error bars because it has many more galaxies than the eBOSS LRG set. Compared to the expected $\alpha=1$, eBOSS agrees in 1$\sigma$, but BOSS is in tension with the expected result. Most points agree in 2$\sigma$ with $\alpha=1$, but they vary whether they are bigger or smaller than such value. This problematic variation w.r.t. the redshift is not yet understood, it appears with the same surveys with the same method. It has been shown in 3D and 2D analysis by \cite{zhao2021completed}, \cite{carvalho2016baryon}, \cite{carvalho2020transverse}, \cite{adame2024desi}, \cite{ferreira2024angular}.

\section{Summary}\label{sec:conclusion}

We discussed the basics of describing and finding the BAO feature from galaxy surveys. Depending on the observation, there are better tools to constrain this probe. When the galaxy's redshift is photometric it is ideal to use the transverse part of the BAO radius because it is not dependent on the line of sight.

Spectroscopic surveys, on the other hand, are the key to studying the BAO in 3D. The problem is the amount of objects observed, because it requires a higher exposure time than the photometric observations, we end up with fewer galaxies within a smaller redshift range. 

In this study, we used two spectroscopic datasets to find the best-fit model to $P(k)$. We used \cite{anderson2012clustering} description of $P(k)$ to find the best-fit model. Here, we are using a different sub-sample compared to the original publication, we got BOSS's $0.3<z<0.65$ instead of slicing in many bins. This was the same sample chosen by \cite{marra2019} which computed the correlation function in redshift space. The other sample $0.6<z<1.0$, eBOSS, had fewer galaxies. We changed the number of polynomial parameters since it has a higher $z$ range. 

Finally, we compared the dilation scale parameter $\alpha$ with results from other surveys. The result from BOSS's fit had smaller error bars than eBOSS's because of the number of LRGs in the sample. WiggleZ and DES Y6 also present big error bars, but the reason is due to the type of observation, photometric surveys are less precise. However, DESI III results match ours in $1\sigma$ and most of their bins are close to unit. 

The modulation of $\alpha$ w.r.t. $z$ is not yet understood by the community. It appears both in three-dimensional and angular tomographic analysis. Whether or not the reason is statistical, we require more bins by different surveys to understand such behaviour. 

\section*{Acknowledgements}
PSF thanks Brazilian funding agency CNPq for PhD scholarship GD 140580/2021-2. RRRR thanks CNPq for partial financial support (grant no. $309868/2021-1$).

Funding for the Sloan Digital Sky Survey IV has been provided by the Alfred P. Sloan Foundation, the U.S. Department of Energy Office of Science, and the Participating Institutions. SDSS acknowledges support and resources from the Center for High-Performance Computing at the University of Utah. The SDSS website is www.sdss.org.

SDSS is managed by the Astrophysical Research Consortium for the Participating Institutions of the SDSS Collaboration including the Brazilian Participation Group, the Carnegie Institution for Science, Carnegie Mellon University, Center for Astrophysics | Harvard \& Smithsonian (CfA), the Chilean Participation Group, the French Participation Group, Instituto de Astrofísica de Canarias, The Johns Hopkins University, Kavli Institute for the Physics and Mathematics of the Universe (IPMU) / University of Tokyo, the Korean Participation Group, Lawrence Berkeley National Laboratory, Leibniz Institut für Astrophysik Potsdam (AIP), Max-Planck-Institut für Astronomie (MPIA Heidelberg), Max-Planck-Institut für Astrophysik (MPA Garching), Max-Planck-Institut für Extraterrestrische Physik (MPE), National Astronomical Observatories of China, New Mexico State University, New York University, University of Notre Dame, Observatório Nacional / MCTI, The Ohio State University, Pennsylvania State University, Shanghai Astronomical Observatory, United Kingdom Participation Group, Universidad Nacional Autónoma de México, University of Arizona, University of Colorado Boulder, University of Oxford, University of Portsmouth, University of Utah, University of Virginia, University of Washington, University of Wisconsin, Vanderbilt University, and Yale University.

The massive production of all MultiDark-Patchy mocks for the BOSS Final Data Release has been performed at the BSC Marenostrum supercomputer, the Hydra cluster at the Instituto de Física Teorica UAM/CSIC, and NERSC at the Lawrence Berkeley National Laboratory. We acknowledge support from the Spanish MICINNs Consolider-Ingenio 2010 Programme under grant MultiDark CSD2009-00064, MINECO Centro de Excelencia Severo Ochoa Programme under grant SEV- 2012-0249, and grant AYA2014-60641-C2-1-P. The MultiDark-Patchy mocks was an effort led from the IFT UAM-CSIC by F. Prada’s group (C.-H. Chuang, S. Rodriguez-Torres and C. Scoccola) in collaboration with C. Zhao (Tsinghua U.), F.-S. Kitaura (AIP), A. Klypin (NMSU), G. Yepes (UAM), and the BOSS galaxy clustering working group.

\printbibliography

@ARTICLE{marra2019,
       author = {{Marra}, Valerio and {Isidro}, Eddy G. C.},
        title = "{A first model-independent radial BAO constraint from the final BOSS sample}",
      journal = {Monthly Notices of the Royal Astronomical Society},
         year = 2019,
        month = aug,
       volume = {487},
       number = {3},
        pages = {3419-3426},
          doi = {10.1093/mnras/stz1557},
archivePrefix = {arXiv},
       eprint = {1808.10695},
 primaryClass = {astro-ph.CO},
       adsurl = {https://ui.adsabs.harvard.edu/abs/2019MNRAS.487.3419M}
}

@ARTICLE{landy1993,
       author = {{Landy}, Stephen D. and {Szalay}, Alexander S.},
        title = "{Bias and Variance of Angular Correlation Functions}",
      journal = {The Astrophysical Journal},
         year = 1993,
        month = jul,
       volume = {412},
        pages = {64},
          doi = {10.1086/172900},
       adsurl = {https://ui.adsabs.harvard.edu/abs/1993ApJ...412...64L}
}

@BOOK{amendola2010,
       author = {{Amendola}, Luca and {Tsujikawa}, Shinji},
        title = "{Dark Energy: Theory and Observations}",
         year = 2010,
         publisher={Cambridge University Press},
       adsurl = {https://ui.adsabs.harvard.edu/abs/2010deto.book.....A}
}

@article{kitaura2016clustering,
    author = {Kitaura, Francisco-Shu and Rodríguez-Torres, Sergio and Chuang, Chia-Hsun and Zhao, Cheng and Prada, Francisco and Gil-Marín, Héctor and Guo, Hong and Yepes, Gustavo and Klypin, Anatoly and Scóccola, Claudia G. and Tinker, Jeremy and McBride, Cameron and Reid, Beth and Sánchez, Ariel G. and Salazar-Albornoz, Salvador and Grieb, Jan Niklas and Vargas-Magana, Mariana and Cuesta, Antonio J. and Neyrinck, Mark and Beutler, Florian and Comparat, Johan and Percival, Will J. and Ross, Ashley},
    title = "{The clustering of galaxies in the SDSS-III Baryon Oscillation Spectroscopic Survey: mock galaxy catalogues for the BOSS Final Data Release}",
    journal = {Monthly Notices of the Royal Astronomical Society},
    volume = {456},
    number = {4},
    pages = {4156-4173},
    year = {2016},
    month = {01},
    issn = {0035-8711},
    doi = {10.1093/mnras/stv2826},
    url = {https://doi.org/10.1093/mnras/stv2826},
    eprint = {https://academic.oup.com/mnras/article-pdf/456/4/4156/9379188/stv2826.pdf},
}

@article{reid2016sdss,
  title={SDSS-III Baryon Oscillation Spectroscopic Survey Data Release 12: galaxy target selection and large-scale structure catalogues},
  author={Reid, Beth and Ho, Shirley and Padmanabhan, Nikhil and Percival, Will J and Tinker, Jeremy and Tojeiro, Rita and White, Martin and Eisenstein, Daniel J and Maraston, Claudia and Ross, Ashley J and others},
  journal={Monthly Notices of the Royal Astronomical Society},
  volume={455},
  number={2},
  pages={1553--1573},
  year={2016},
  publisher={Oxford University Press}
}

@article{feldman1993power,
    author = {Feldman, Hume A. and Kaiser, Nick and Peacock, John A.},
    title = {Power-Spectrum Analysis of Three-dimensional Redshift Surveys},
    journal = {The Astrophysical Journal},
    year = 1994,
    month = may,
    volume = {426},
    pages = {23},
    doi = {10.1086/174036},}

@article{guth81,
  title = {Inflationary universe: A possible solution to the horizon and flatness problems},
  author = {Guth, Alan H.},
  journal = {Phys. Rev. D},
  volume = {23},
  issue = {2},
  pages = {347--356},
  numpages = {0},
  year = {1981},
  publisher = {American Physical Society},
  doi = {10.1103/PhysRevD.23.347},
  url = {https://link.aps.org/doi/10.1103/PhysRevD.23.347}
}

@article{collaboration2018planck,
  title={Planck 2018 results. X. Constraints on inflation},
  author={{Planck Collaboration 2018}},
  year={2018}
}

@article{ade2016planck,
  title={Planck 2015 results-xiii. cosmological parameters},
  author={{Planck Collaboration 2015}},
  journal={Astronomy \& Astrophysics},
  volume={594},
  pages={A13},
  year={2016},
  publisher={EDP sciences}
}

@article{eisenstein2005detection,
  title={Detection of the baryon acoustic peak in the large-scale correlation function of SDSS luminous red galaxies},
  author={Eisenstein and Zehavi, Idit and Hogg, David W and Scoccimarro, Roman and Blanton, Michael R and Nichol, Robert C and Scranton, Ryan and Seo, Hee-Jong and Tegmark, Max and Zheng, Zheng and others},
  journal={The Astrophysical Journal},
  volume={633},
  number={2},
  pages={560},
  year={2005},
  publisher={IOP Publishing}
}

@article{sanchez2013precise,
  title={Precise measurement of the radial baryon acoustic oscillation scales in galaxy redshift surveys},
  author={Sánchez, E and Alonso, D and Sánchez, FJ and Garc{\'\i}a-Bellido, J and Sevilla, I},
  journal={Monthly Notices of the Royal Astronomical Society},
  volume={434},
  number={3},
  pages={2008--2019},
  year={2013},
  publisher={The Royal Astronomical Society}
}

@article{hand2018nbodykit,
  title={nbodykit: An open-source, massively parallel toolkit for large-scale structure},
  author={Hand, Nick and Feng, Yu and Beutler, Florian and Li, Yin and Modi, Chirag and Seljak, Uro{\v{s}} and Slepian, Zachary},
  journal={The Astronomical Journal},
  volume={156},
  number={4},
  pages={160},
  year={2018},
  publisher={IOP Publishing}
}

@article{eisenstein1998baryonic,
  title={Baryonic features in the matter transfer function},
  author={Eisenstein, Daniel J and Hu, Wayne},
  journal={The Astrophysical Journal},
  volume={496},
  number={2},
  pages={605},
  year={1998},
  publisher={IOP Publishing}
}

@article{rodriguez2016clustering,
  title={The clustering of galaxies in the SDSS-III Baryon Oscillation Spectroscopic Survey: modelling the clustering and halo occupation distribution of BOSS CMASS galaxies in the Final Data Release},
  author={Rodríguez-Torres, Sergio A and Chuang, Chia-Hsun and Prada, Francisco and Guo, Hong and Klypin, Anatoly and Behroozi, Peter and Hahn, Chang Hoon and Comparat, Johan and Yepes, Gustavo and Montero-Dorta, Antonio D and others},
  journal={Monthly Notices of the Royal Astronomical Society},
  volume={460},
  number={2},
  pages={1173--1187},
  year={2016},
  publisher={Oxford University Press}
}

@article{anderson2014clustering,
  title={The clustering of galaxies in the SDSS-III Baryon Oscillation Spectroscopic Survey: baryon acoustic oscillations in the Data Releases 10 and 11 Galaxy samples},
  author={Anderson, Lauren and Aubourg, {\'E}ric and Bailey, Stephen and Beutler, Florian and Bhardwaj, Vaishali and Blanton, Michael and Bolton, Adam S and Brinkmann, Jon and Brownstein, Joel R and Burden, Angela and others},
  journal={Monthly Notices of the Royal Astronomical Society},
  volume={441},
  number={1},
  pages={24--62},
  year={2014},
  publisher={Oxford University Press}
}

@article{zhao2016clustering,
  title={The clustering of galaxies in the completed SDSS-III Baryon Oscillation Spectroscopic Survey: tomographic BAO analysis of DR12 combined sample in Fourier space},
  author={Zhao, Gong-Bo and Wang, Yuting and Saito, Shun and Wang, Dandan and Ross, Ashley J and Beutler, Florian and Grieb, Jan Niklas and Chuang, Chia-Hsun and Kitaura, Francisco-Shu and Rodriguez-Torres, Sergio and others},
  journal={Monthly Notices of the Royal Astronomical Society},
  volume={466},
  number={1},
  pages={762--779},
  year={2016},
  publisher={Oxford University Press}
}

@article{foreman2013emcee,
  title={emcee: the MCMC hammer},
  author={Foreman-Mackey, Daniel and Hogg, David W and Lang, Dustin and Goodman, Jonathan},
  journal={Publications of the Astronomical Society of the Pacific},
  volume={125},
  number={925},
  pages={306},
  year={2013},
  publisher={IOP Publishing}
}

@article{tegmark2006cosmological,
  title={Cosmological constraints from the SDSS luminous red galaxies},
  author={Tegmark, Max and Eisenstein, Daniel J and Strauss, Michael A and Weinberg, David H and Blanton, Michael R and Frieman, Joshua A and Fukugita, Masataka and Gunn, James E and Hamilton, Andrew JS and Knapp, Gillian R and others},
  journal={Physical Review D},
  volume={74},
  number={12},
  pages={123507},
  year={2006},
  publisher={APS}
}

@article{pearcescikit,
  author={Hans Dembinski and Piti Ongmongkolkul et al.},
  title={scikit-hep/iminuit},
  DOI={10.5281/zenodo.4310361},
  publisher={Zenodo},
  year={2020},
  month={12},
  url={https://doi.org/10.5281/zenodo.4310361}
}

@article{hinshaw2007three,
  title={Three-year wilkinson microwave anisotropy probe (wmap*) observations: Temperature analysis},
  author={Hinshaw, G and Nolta, MR and Bennett, CL and Bean, R and Dor{\'e}, O and Greason, MR and Halpern, M and Hill, RS and Jarosik, N and Kogut, A and others},
  journal={The Astrophysical Journal Supplement Series},
  volume={170},
  number={2},
  pages={288},
  year={2007},
  publisher={IOP Publishing}
}

@article{ross2017clustering,
  title={The clustering of galaxies in the completed SDSS-III Baryon Oscillation Spectroscopic Survey: Observational systematics and baryon acoustic oscillations in the correlation function},
  author={Ross, Ashley J and Beutler, Florian and Chuang, Chia-Hsun and Pellejero-Ibanez, Marcos and Seo, Hee-Jong and Vargas-Magana, Mariana and Cuesta, Antonio J and Percival, Will J and Burden, Angela and Sánchez, Ariel G and others},
  journal={Monthly Notices of the Royal Astronomical Society},
  volume={464},
  number={1},
  pages={1168--1191},
  year={2017},
  publisher={Oxford University Press}
}

@article{vargas2018clustering,
  title={The clustering of galaxies in the completed SDSS-III Baryon Oscillation Spectroscopic Survey: theoretical systematics and Baryon Acoustic Oscillations in the galaxy correlation function},
  author={Vargas-Magaña, Mariana and Ho, Shirley and Cuesta, Antonio J and O'Connell, Ross and Ross, Ashley J and Eisenstein, Daniel J and Percival, Will J and Grieb, Jan Niklas and Sánchez, Ariel G and Tinker, Jeremy L and others},
  journal={Monthly Notices of the Royal Astronomical Society},
  volume={477},
  number={1},
  pages={1153--1188},
  year={2018},
  publisher={Oxford University Press}
}

@article{beutler2017clustering,
  title={The clustering of galaxies in the completed SDSS-III Baryon Oscillation Spectroscopic Survey: anisotropic galaxy clustering in Fourier space},
  author={Beutler, Florian and Seo, Hee-Jong and Saito, Shun and Chuang, Chia-Hsun and Cuesta, Antonio J and Eisenstein, Daniel J and Gil-Mar{\'\i}n, H{\'e}ctor and Grieb, Jan Niklas and Hand, Nick and Kitaura, Francisco-Shu and others},
  journal={Monthly Notices of the Royal Astronomical Society},
  volume={466},
  number={2},
  pages={2242--2260},
  year={2017},
  publisher={Oxford University Press}
}

@article{anderson2012clustering,
  title={The clustering of galaxies in the SDSS-III Baryon Oscillation Spectroscopic Survey: baryon acoustic oscillations in the Data Release 9 spectroscopic galaxy sample},
  author={Anderson, Lauren and Aubourg, Eric and Bailey, Stephen and Bizyaev, Dmitry and Blanton, Michael and Bolton, Adam S and Brinkmann, Jon and Brownstein, Joel R and Burden, Angela and Cuesta, Antonio J and others},
  journal={Monthly Notices of the Royal Astronomical Society},
  volume={427},
  number={4},
  pages={3435--3467},
  year={2012},
  publisher={Blackwell Science Ltd Oxford, UK}
}

@article{kaiser1984spatial,
    author = "Kaiser, Nick",
    title = "{On the Spatial correlations of Abell clusters}",
    doi = "10.1086/184341",
    journal = "Astrophys. J. Lett.",
    volume = "284",
    pages = "L9--L12",
    year = "1984"
}

@article{peebles1970primeval,
  title={Primeval adiabatic perturbation in an expanding universe},
  author={Peebles, Philip JE and Yu, JT},
  journal={The Astrophysical Journal},
  volume={162},
  pages={815},
  year={1970}
}

@article{sunyaev1970small,
  title={Small-scale fluctuations of relic radiation},
  author={Sunyaev, Rashid A and Zeldovich, Ya B},
  journal={Astrophysics and Space Science},
  volume={7},
  number={1},
  pages={3--19},
  year={1970},
  publisher={Springer}
}

@article{peebles1973statistical,
  title={Statistical analysis of catalogs of extragalactic objects. I. Theory},
  author={Peebles,P. J. E.},
  journal={The Astrophysical Journal},
  volume={185},
  pages={413--440},
  year={1973}
}

@article{baumann2009tasi,
  title={TASI lectures on inflation},
  author={Baumann, Daniel},
  journal={arXiv preprint arXiv:0907.5424},
  year={2009}
}

@article{bagla2005cosmological,
  title={Cosmological N-body simulation: Techniques, scope and status},
  author={Bagla, Jasjeet Singh},
  journal={Current science},
  pages={1088--1100},
  year={2005},
  publisher={JSTOR}
}

@article{springel2005simulations,
  title={Simulations of the formation, evolution and clustering of galaxies and quasars},
  author={Springel, Volker and White, Simon DM and Jenkins, Adrian and Frenk, Carlos S and Yoshida, Naoki and Gao, Liang and Navarro, Julio and Thacker, Robert and Croton, Darren and Helly, John and others},
  journal={nature},
  volume={435},
  number={7042},
  pages={629--636},
  year={2005},
  publisher={Nature Publishing Group}
}

@article{cole1998mock,
  title={Mock 2dF and SDSS galaxy redshift surveys},
  author={Cole, Shaun and Hatton, Steve and Weinberg, David H and Frenk, Carlos S},
  journal={Monthly Notices of the Royal Astronomical Society},
  volume={300},
  number={4},
  pages={945--966},
  year={1998},
  publisher={Blackwell Science Ltd Oxford, UK}
}

@article{klypin2016multidark,
  title={MultiDark simulations: the story of dark matter halo concentrations and density profiles},
  author={Klypin, Anatoly and Yepes, Gustavo and Gottl{\"o}ber, Stefan and Prada, Francisco and Hess, Steffen},
  journal={Monthly Notices of the Royal Astronomical Society},
  volume={457},
  number={4},
  pages={4340--4359},
  year={2016},
  publisher={The Royal Astronomical Society}
}

@article{zel1970gravitational,
  title={Gravitational instability: An approximate theory for large density perturbations.},
  author={Zel'Dovich, Ya B},
  journal={Astronomy and astrophysics},
  volume={5},
  pages={84--89},
  year={1970}
}

@article{zhao2021completed,
  title={The completed SDSS-IV extended Baryon Oscillation Spectroscopic Survey: 1000 multi-tracer mock catalogues with redshift evolution and systematics for galaxies and quasars of the final data release},
  author={Zhao, Cheng and Chuang, Chia-Hsun and Bautista, Julian and De Mattia, Arnaud and Raichoor, Anand and Ross, Ashley J and Hou, Jiamin and Neveux, Richard and Tao, Charling and Burtin, Etienne and others},
  journal={Monthly Notices of the Royal Astronomical Society},
  volume={503},
  number={1},
  pages={1149--1173},
  year={2021},
  publisher={Oxford University Press}
}

@article{eisenstein2005detection1,
  title={Detection of the baryon acoustic peak in the large-scale correlation function of SDSS luminous red galaxies},
  author={Eisenstein, Daniel J and Zehavi, Idit and Hogg, David W and Scoccimarro, Roman and Blanton, Michael R and Nichol, Robert C and Scranton, Ryan and Seo, Hee-Jong and Tegmark, Max and Zheng, Zheng and others},
  journal={The Astrophysical Journal},
  volume={633},
  number={2},
  pages={560},
  year={2005},
  publisher={IOP Publishing}
}

@article{cole20052df,
  title={The 2dF Galaxy Redshift Survey: power-spectrum analysis of the final data set and cosmological implications},
  author={Cole, Shaun and Percival, Will J and Peacock, John A and Norberg, Peder and Baugh, Carlton M and Frenk, Carlos S and Baldry, Ivan and Bland-Hawthorn, Joss and Bridges, Terry and Cannon, Russell and others},
  journal={Monthly Notices of the Royal Astronomical Society},
  volume={362},
  number={2},
  pages={505--534},
  year={2005},
  publisher={The Royal Astronomical Society}
}

@article{accetta2022seventeenth,
  title={The Seventeenth Data Release of the Sloan Digital Sky Surveys: Complete Release of MaNGA, MaStar, and APOGEE-2 Data},
  author={Accetta, Katherine and Aerts, Conny and Aguirre, V{\'\i}ctor Silva and Ahumada, Romina and Ajgaonkar, Nikhil and Ak, N Filiz and Alam, Shadab and Prieto, Carlos Allende and Almeida, Andr{\'e}s and Anders, Friedrich and others},
  journal={The Astrophysical Journal Supplement Series},
  volume={259},
  number={2},
  pages={35},
  year={2022},
  publisher={IOP Publishing}
}

@article{drinkwater2010wigglez,
  title={The WiggleZ Dark Energy Survey: survey design and first data release},
  author={Drinkwater, Michael J and Jurek, Russell J and Blake, Chris and Woods, David and Pimbblet, Kevin A and Glazebrook, Karl and Sharp, Rob and Pracy, Michael B and Brough, Sarah and Colless, Matthew and others},
  journal={Monthly Notices of the Royal Astronomical Society},
  volume={401},
  number={3},
  pages={1429--1452},
  year={2010},
  publisher={Blackwell Publishing Ltd Oxford, UK}
}

@article{blake2012wigglez,
  title={The WiggleZ Dark Energy Survey: Joint measurements of the expansion and growth history at $z< 1$},
  author={Blake, Chris and Brough, Sarah and Colless, Matthew and Contreras, Carlos and Couch, Warrick and Croom, Scott and Croton, Darren and Davis, Tamara M and Drinkwater, Michael J and Forster, Karl and others},
  journal={Monthly Notices of the Royal Astronomical Society},
  volume={425},
  number={1},
  pages={405--414},
  year={2012},
  publisher={Blackwell Science Ltd Oxford, UK}
}

@article{kazin2014wigglez,
  title={The WiggleZ Dark Energy Survey: improved distance measurements to $z= 1$ with reconstruction of the baryonic acoustic feature},
  author={Kazin, Eyal A and Koda, Jun and Blake, Chris and Padmanabhan, Nikhil and Brough, Sarah and Colless, Matthew and Contreras, Carlos and Couch, Warrick and Croom, Scott and Croton, Darren J and others},
  journal={Monthly Notices of the Royal Astronomical Society},
  volume={441},
  number={4},
  pages={3524--3542},
  year={2014},
  publisher={Oxford University Press}
}

@article{crocce2019dark,
  title={Dark Energy Survey year 1 results: galaxy sample for BAO measurement},
  author={Crocce, Martin and Ross, AJ and Sevilla-Noarbe, Ignacio and Gaztanaga, Enrique and Elvin-Poole, Jack and Avila, Santiago and Alarcon, Alex and Chan, Kwan Chuen and Banik, N and Carretero, Jorge and others},
  journal={Monthly Notices of the Royal Astronomical Society},
  volume={482},
  number={2},
  pages={2807--2822},
  year={2019},
  publisher={Oxford University Press}
}

@article{salvato2019many,
  title={The many flavours of photometric redshifts},
  author={Salvato, Mara and Ilbert, Olivier and Hoyle, Ben},
  journal={Nature Astronomy},
  volume={3},
  number={3},
  pages={212--222},
  year={2019},
  publisher={Nature Publishing Group}
}

@article{DESY3,
    author = {Rosell, A Carnero and Rodriguez-Monroy, M and Crocce, M and Elvin-Poole, J and Porredon, A and Ferrero, I and Mena-Fernández, J and Cawthon, R and De Vicente, J and Gaztanaga, E and Ross, A J and Sanchez, E and Sevilla-Noarbe, I and Alves, O and Andrade-Oliveira, F and Asorey, J and Avila, S and Brandao-Souza, A and Camacho, H and Chan, K C and Ferté, A and Muir, J and Riquelme, W and Rosenfeld, R and Cid, D Sanchez and Hartley, W G and Weaverdyck, N and Abbott, T and Aguena, M and Allam, S and Annis, J and Bertin, E and Brooks, D and Buckley-Geer, E and Burke, D and Calcino, J and Carollo, D and Kind, M Carrasco and Carretero, J and Castander, F and Choi, A and Conselice, C and Costanzi, M and da Costa, L and da Silva Pereira, M E and Davis, T and Desai, S and Diehl, H T and Doel, P and Drlica-Wagner, A and Eckert, K and Everett, S and Evrard, A and Flaugher, B and Fosalba, P and Frieman, J and Garcia-Bellido, J and Gerdes, D and Giannantonio, T and Glazebrook, K and Gruen, D and Gruendl, R and Gschwend, J and Gutierrez, G and Hinton, S and Hollowood, D and Honscheid, K and Hoyle, B and Huterer, D and James, D and Kim, A and Krause, E and Kuehn, K and Lahav, O and Lewis, G and Lidman, C and Lima, M and Maia, M and Malik, U and Marshall, J and Menanteau, F and Miquel, R and Mohr, J and Moller, A and Morgan, R and Ogando, R and Palmese, A and Paz-Chinchon, F and Percival, W and Pieres, A and Malagón, A Plazas and Roodman, A and Scarpine, V and Schubnell, M and Serrano, S and Sharp, R and Sheldon, E and Smith, M and Soares-Santos, M and Suchyta, E and Swanson, M and Tarle, G and Thomas, D and To, C and Tucker, B and Tucker, D and Uddin, S and Varga, T N and DES Collaboration},
    title = "{Dark Energy Survey Year 3 results: galaxy sample for BAO measurement}",
    journal = {Monthly Notices of the Royal Astronomical Society},
    volume = {509},
    number = {1},
    pages = {778-799},
    year = {2021},
    month = {10},
    issn = {0035-8711},
    doi = {10.1093/mnras/stab2995},
    url = {https://doi.org/10.1093/mnras/stab2995},
    eprint = {https://academic.oup.com/mnras/article-pdf/509/1/778/41118809/stab2995.pdf},
}

@article{newman2013deep2,
  title={The deep2 galaxy redshift survey: Design, observations, data reduction, and redshifts},
  author={Newman, Jeffrey A and Cooper, Michael C and Davis, Marc and Faber, SM and Coil, Alison L and Guhathakurta, Puragra and Koo, David C and Phillips, Andrew C and Conroy, Charlie and Dutton, Aaron A and others},
  journal={The Astrophysical Journal Supplement Series},
  volume={208},
  number={1},
  pages={5},
  year={2013},
  publisher={IOP Publishing}
}

@article{le2005vimos,
  title={The VIMOS VLT deep survey-First epoch VVDS-deep survey: 11 564 spectra with 17.5$\leq$ I $\leq$ 24, and the redshift distribution over 0$\leq$ z$\leq$ 5},
  author={Le Fèvre, Olivier and Vettolani, G and Garilli, B and Tresse, L and Bottini, DBVL and Le Brun, V and Maccagni, D and Picat, JP and Scaramella, R and Scodeggio, M and others},
  journal={Astronomy \& Astrophysics},
  volume={439},
  number={3},
  pages={845--862},
  year={2005},
  publisher={EDP Sciences}
}

@article{scodeggio2018vimos,
  title={The VIMOS Public Extragalactic Redshift Survey (VIPERS)-Full spectroscopic data and auxiliary information release (PDR-2)},
  author={Scodeggio, MARCO and Guzzo, L and Garilli, BIANCA and Granett, BR and Bolzonella, M and De La Torre, S and Abbas, U and Adami, C and Arnouts, S and Bottini, D and others},
  journal={Astronomy \& Astrophysics},
  volume={609},
  pages={A84},
  year={2018},
  publisher={EDP Sciences}
}

@article{wang2020clustering,
  title={The clustering of the SDSS-IV extended baryon oscillation spectroscopic survey DR16 luminous red galaxy and emission-line galaxy samples: cosmic distance and structure growth measurements using multiple tracers in configuration space},
  author={Wang, Yuting and Zhao, Gong-Bo and Zhao, Cheng and Philcox, Oliver HE and Alam, Shadab and Tamone, Am{\'e}lie and De Mattia, Arnaud and Ross, Ashley J and Raichoor, Anand and Burtin, Etienne and others},
  journal={Monthly Notices of the Royal Astronomical Society},
  volume={498},
  number={3},
  pages={3470--3483},
  year={2020},
  publisher={Oxford University Press}}

@article{abbott2024dark,
  title={Dark Energy Survey: A 2.1\% measurement of the angular Baryonic Acoustic Oscillation scale at redshift $ z_{eff} $= 0.85 from the final dataset},
  author={Abbott, TMC and Adamow, M and Aguena, M and Allam, S and Alves, O and Amon, A and Andrade-Oliveira, F and Asorey, J and Avila, S and Bacon, D and others},
  journal={arXiv preprint arXiv:2402.10696},
  year={2024}
}

@article{sanchez2011tracing,
  title={Tracing the sound horizon scale with photometric redshift surveys},
  author={S{\'a}nchez, Eusebio and Carnero, Aurelio and Garc{\'\i}a-Bellido, Juan and Gaztanaga, Enrique and De Simoni, F and Crocce, Mart{\i}n and Cabr{\'e}, Anna and Fosalba, Pablo and Alonso, David},
  journal={Monthly Notices of the Royal Astronomical Society},
  volume={411},
  number={1},
  pages={277--288},
  year={2011},
  publisher={Blackwell Publishin6790g Ltd Oxford, UK}
}

@article{ferreira2024angular,
  title={Angular correlation function from sample covariance with BOSS and eBOSS LRG},
  author={Ferreira, Paula S and Reis, Ribamar RR},
  journal={The European Physical Journal C},
  volume={84},
  number={5},
  pages={466},
  year={2024},
  publisher={Springer}
}

@article{menote2022baryon,
  title={Baryon acoustic oscillations in thin redshift shells from BOSS DR12 and eBOSS DR16 galaxies},
  author={Menote, Ranier and Marra, Valerio},
  journal={Monthly Notices of the Royal Astronomical Society},
  volume={513},
  number={2},
  pages={1600--1608},
  year={2022},
  publisher={Oxford University Press}
}

@article{adame2024desi,
  title={DESI 2024 III: Baryon Acoustic Oscillations from Galaxies and Quasars},
  author={Adame, AG and Aguilar, J and Ahlen, S and Alam, S and Alexander, DM and Alvarez, M and Alves, O and Anand, A and Andrade, U and Armengaud, E and others},
  journal={arXiv preprint arXiv:2404.03000},
  year={2024}
}

@article{carvalho2016baryon,
  title={Baryon acoustic oscillations from the SDSS DR10 galaxies angular correlation function},
  author={Carvalho, GC and Bernui, A and Benetti, M and Carvalho, JC and Alcaniz, JS},
  journal={Physical Review D},
  volume={93},
  number={2},
  pages={023530},
  year={2016},
  publisher={APS}
}

@article{carvalho2020transverse,
  title={The transverse baryonic acoustic scale from the SDSS DR11 galaxies},
  author={Carvalho, GC and Bernui, A and Benetti, M and Carvalho, JC and de Carvalho, E and Alcaniz, JS},
  journal={Astroparticle Physics},
  volume={119},
  pages={102432},
  year={2020},
  publisher={Elsevier}
}

@article{perez2024fiducial,
  title={Fiducial-Cosmology-dependent systematics for the DESI 2024 BAO Analysis},
  author={P{\'e}rez-Fern{\'a}ndez, A and Medina-Varela, L and Ruggeri, R and Vargas-Maga{\~n}a, M and Seo, H and Padmanabhan, N and Ishak, M and Aguilar, J and Ahlen, S and Alam, S and others},
  journal={arXiv preprint arXiv:2406.06085},
  year={2024}
}

\end{document}